\documentclass{emulateapj}

\usepackage{wrapfig}
\usepackage{amsmath}
\usepackage{apjfonts}

\newcommand{\tnm}{\tablenotemark}
\newcommand{\tnt}{\tablenotetext}

\newcommand{\fnm}{\footnotemark}
\newcommand{\fnt}{\footnotetext}

\newcommand{\ergs}{erg s$^{-1}$}

\newcommand{\pers}{s$^{-1}$}
\newcommand{\cdens}{cm$^{-2}$}

\newcommand{\chandra}{{\it Chandra}}

\newcommand{\average}[1]{\ensuremath{\langle#1\rangle} }

\newcommand{\msun}{M_{\sun}}

\newcommand{\bootes}{Bo\"{o}tes} 
 
\newcommand{\spitzer}{{\it Spitzer}}

\newcommand{\hminus}{$h^{-1}$}
\newcommand{\hmpc}{$h^{-1}$ Mpc}
\newcommand{\hmsun}{$h^{-1}$ $M_{\sun}$}

\newcommand{\qsoone}{QSO-1}
\newcommand{\qsotwo}{Obs-QSO}

\begin{document}

\slugcomment{Accepted for publication in The Astrophysical Journal}

\title{Clustering of obscured and unobscured quasars in the Bo\"{o}tes field:  \\ Placing rapidly growing black holes in the cosmic web}

\shorttitle{CLUSTERING OF OBSCURED AND UNOBSCURED QUASARS}
\shortauthors{HICKOX ET AL.}
\author{Ryan C. Hickox\altaffilmark{1,2,3}}
\author{Adam D. Myers\altaffilmark{4}}
\author{Mark Brodwin\altaffilmark{3}}
\author{David M. Alexander\altaffilmark{1}}
\author{William R. Forman\altaffilmark{3}}
\author{Christine Jones\altaffilmark{3}}
\author{Stephen S. Murray\altaffilmark{3,5}}
\author{Michael J.~I. Brown\altaffilmark{6}}
\author{Richard J. Cool\altaffilmark{7,8}}
\author{Christopher S. Kochanek\altaffilmark{9}}
\author{Arjun Dey\altaffilmark{10}}
\author{Buell T. Jannuzi\altaffilmark{10}}
\author{Daniel Eisenstein\altaffilmark{3,11}}
\author{Roberto J. Assef\altaffilmark{12,13}}
\author{Peter R. Eisenhardt\altaffilmark{12}}
\author{Varoujan Gorjian\altaffilmark{12}}
\author{Daniel Stern\altaffilmark{12}}
\author{Emeric Le Floc'h\altaffilmark{14}}
\author{Nelson Caldwell\altaffilmark{3}}
\author{Andrew D. Goulding\altaffilmark{1,3}}
\author{James R. Mullaney\altaffilmark{1,14}}

\altaffiltext{1}{Department of Physics, Durham University, South Road, Durham, DH1 3LE, United Kingdom; ryan.hickox@durham.ac.uk.}
\altaffiltext{2}{STFC Postdoctoral Fellow.}
\altaffiltext{3}{Harvard-Smithsonian Center for Astrophysics, 60 Garden Street,
 Cambridge, MA 02138.}
\altaffiltext{4}{Department of Astronomy, University of Illinois at Urbana-Champaign, Urbana IL 61801.}
\altaffiltext{5}{Department of Physics \& Astronomy, The Johns Hopkins University, 3400 N.\ Charles Street, Baltimore, MD 21218.}
\altaffiltext{6}{School of Physics, Monash University, Clayton 3800, Victoria, Australia.}
\altaffiltext{7}{Princeton University Observatory, Peyton Hall, Princeton, NJ  08544.}
\altaffiltext{8}{Hubble Fellow and Carnegie-Princeton Fellow.}
\altaffiltext{9}{Department of Astronomy, The Ohio State University, 140 West 18th Avenue, Columbus, OH 43210.}
\altaffiltext{10}{National Optical Astronomy Observatory, Tucson, AZ  85726.}
\altaffiltext{11}{Steward Observatory, 933 North Cherry Avenue, Tucson, AZ 85721.}
\altaffiltext{12}{Jet Propulsion Laboratory, California Institute of Technology, Pasadena, CA 91109.}
\altaffiltext{13}{NASA Postdoctoral Program Fellow.}
\altaffiltext{14}{Laboratoire AIM-Paris-Saclay, CEA/DSM/Irfu - CNRS - Universit\'{e} Paris Diderot, CE-Saclay, pt courrier 131, 91191 Gif-sur-Yvette, France.}


\begin{abstract}

We present the first measurement of the spatial clustering of
mid-infrared selected obscured and unobscured quasars, using a sample
in the redshift range $0.7 < z < 1.8$ selected from the 9 deg$^2$
\bootes\ multiwavelength survey.  Recently the {\em Spitzer Space
Telescope} and X-ray observations have revealed large populations of
obscured quasars that have been inferred from models of the  X-ray
background and supermassive black hole evolution.  To date,
little is known about obscured quasar clustering, which allows us to
measure the masses of their host dark matter halos and explore their
role in the cosmic evolution of black holes and galaxies.  In this
study we use a sample of 806 mid-infrared selected quasars and
$\approx$250,000 galaxies to calculate the projected quasar-galaxy
cross-correlation function $w_p(R)$.  The observed clustering yields
characteristic dark matter halo masses of $\log(M_{\rm halo} [h^{-1}
M_{\sun}]) =12.7^{+0.4}_{-0.6}$ and $13.3^{+0.3}_{-0.4}$ for
unobscured quasars ({\qsoone}s) and obscured quasars ({\qsotwo}s),
respectively.  The results for {\qsoone}s are in excellent agreement with
previous measurements for optically-selected quasars, while we
conclude that the {\qsotwo}s are {\em at least} as
strongly clustered as the {\qsoone}s.  We test for the effects of
photometric redshift errors on the optically-faint {\qsotwo}s, and find
that our method yields a robust lower limit on the clustering;
photo-$z$ errors may cause us to underestimate the clustering
amplitude of the {\qsotwo}s by at most $\sim$20\%.  We compare our results
to previous studies, and speculate on physical implications of
stronger clustering for obscured quasars.
\end{abstract}


\section{Introduction}
\label{intro}

Supermassive black holes with masses $\gtrsim 10^6 \msun$ are
ubiquitous in the nuclei of local galaxies of moderate to high mass
\citep[e.g.,][]{korm95}.  It is now well established that most of the
total mass in black holes in the nearby Universe was accreted in luminous
episodes with high Eddington rates \citep[e.g.,][]{solt82, yu02smbh},
with the growth for massive ($M_{\rm BH}\gtrsim10^8 \msun$) black
holes occurring predominantly at $z\gtrsim 1$
\citep[e.g.,][]{merl08agnsynth, shan09agnbh}.  These rapidly accreting
black holes are most readily identified as bright optical quasars with
characteristic broad ($>1000$ km \pers) emission lines, and luminous
continuum emission that can dominate the light from the host galaxy,
particularly at ultraviolet and optical wavelengths
\citep[e.g.,][]{elvi94,rich06, schn07sdssqso}.  Optical quasars thus
provide powerful tools for tracing the rapid growth of black holes
over cosmic time \citep[e.g.,][]{croo04twodfqz, rich05, fan06}.

However, it is increasingly clear that a significant fraction of the
quasar population does not show characteristic blue continua or broad
lines because their nuclear emission regions are obscured.  Key
evidence for the existence of obscured (Type 2) quasars comes from
synthesis models of the cosmic X-ray background
\citep[e.g.,][]{coma95, gill07cxb}, as well as direct identification of these
objects through various observational techniques.  These include selection of
luminous quasars with only narrow optical lines \citep{zaka03,zaka04,
  zaka05, reye08qso2} or relatively weak X-ray emission
\citep{ptak06qso2, vign06qso2, vign10qso2}, detection of powerful
radio galaxies lacking strong nuclear optical continua or broad lines
\citep[e.g.,][]{mcca93highzradio, seym07radiohosts}, and detection
of X-ray sources that are optically faint
\citep[e.g.,][]{alex01xfaint, ster02, trei04, main05xfaint}, have hard X-ray
spectra \citep[e.g.,][]{vign09qso2}, or have radio bright, optically
weak counterparts \citep[e.g.,][]{mart06}.

\defcitealias{hick07abs}{H07}

With the launch of the {\em Spitzer Space Telescope}, large numbers of
obscured quasars can now be efficiently identified based on their
characteristic (roughly power-law) spectral energy distributions
(SEDs) at mid-infrared (mid-IR) wavelengths ($\approx$3--24 \micron).
Because mid-IR emission is less strongly affected by dust extinction than
optical and ultraviolet light, obscured quasars can appear
similar to their unobscured counterparts in the mid-IR, but have
optical emission characteristic of their host galaxies.  A number of
studies using mid-IR colors \citep[hereafter H07]{lacy04, ster05, rowa05, hick07abs},
SED fitting \citep{alon06, donl07plaw}, or selecting objects based on
similarities to mid-IR quasar templates \citep[e.g.,][]{poll06} have
been successful in identifying large numbers of dust-obscured quasars,
indicating that a large fraction, and possibly a majority of rapid
black hole growth is obscured by dust.

These large new samples enable detailed statistical studies that can
explore the role of obscured quasars in galaxy and black hole
evolution.  At present there are a number of possible physical
scenarios for obscured quasars; in the simplest ``unified models'',
obscuration is attributed to a broadly axisymmetric ``torus'' of dust
that is part of the central engine, so obscuration is entirely an
orientation effect \citep[e.g.,][]{anto93, urry95}.  Alternatively,
obscuration may not be due to a central ``torus'' but to larger dust
structures such as those predicted during major mergers of galaxies
\citep[e.g.,][]{silk98, spri05,hopk06apjs}, and obscured quasars may
represent an early evolutionary phase when the growing black hole
cannot produce a high enough accretion luminosity to expel the
surrounding material \citep[e.g.,][]{hopk08frame1, king10bhmass}.
Observations have revealed evidence for obscuration by a ``torus'' in
some cases and by galactic-scale structures in others
\citep[e.g.,][]{zaka05, page04submm, mart09millqso}, and while there
are examples of obscured quasars that show clear signs of radiative
feedback on interstellar gas, it is unclear
whether they are driving the galaxy-scale outflows invoked in
evolutionary models  \citep{gree11obsqso}. Thus the physical nature of obscured quasars
remains poorly understood, and analyses with large samples of mid-IR
selected quasars will be essential for a more complete understanding
of rapidly growing, obscured black holes.

One particularly powerful observational tool is spatial clustering,
which allows us to measure the masses of the dark matter halos in
which quasars reside.  Clustering studies of unobscured quasars have
shown that the masses of quasar host halos are remarkably constant
with cosmic time, with $M_{\rm halo} \sim 3\times10^{12}$ \hmsun\ over
the large redshift range $0 < z \lesssim 5$
\citep[e.g.,][]{porc04clust,croo05,coil07a, myer07clust1,shen07clust,
  daan08clust,padm09qsored, ross09qsoclust}. This lack of variation in
halo mass implies that the bias factor (clustering relative to the
underlying dark matter) is an increasing function of redshift, since
the dark matter is more weakly clustered earlier in cosmic time.
The characteristic $M_{\rm halo}$ provides a strong constraint on
models of quasar fueling by the major mergers of gas-rich galaxies
\citep[e.g.,][]{kauf00merge, spri05, hopk06merge}, secular
instabilities \citep[e.g.,][]{mo98disk, bowe06gal, genz08ifsz2} or
accretion of recycled cold gas from evolved stars \citep{ciot07flare,
  ciot10flare}, and may be related to quasars' role in regulating star
formation and the emergence of the red galaxy population in halos of
roughly similar mass $\sim$$10^{12}$--$10^{13}$ \hmsun\ (e.g.,
\citealt{coil08galclust, brow08halo, conr09halo, tink10clust}).

Despite the power of clustering measurements in understanding quasar
populations, little is known about the clustering of
obscured quasars.  Some measurements of lower-luminosity AGNs
indicate no significant difference between obscured and unobscured
sources \citep{cons06clust, li06agnclust, gand06xclust,
  mand09agnclust, gill09xclust, hick09corr}.  However, these AGNs
likely have different physical drivers compared to powerful quasars
\citep[e.g.,][]{hopk06low}.  For obscured quasars at high luminosities
($L_{\rm bol} \sim 10^{46}$ \ergs) and high redshift ($z \gtrsim 1$),
the clustering has remained largely unexplored.

In this paper we present the first measurement of the clustering of
mid-IR selected obscured quasars and make direct comparisons to their
unobscured counterparts.  We use a large sample of quasars
(both obscured and unobscured) in the redshift range $0.7 < z < 1.8$
selected on the basis of IRAC colors by \citetalias{hick07abs}, using
data from the 9 deg$^2$ \bootes\ multiwavelength survey.  We also
employ a sample of $\approx$250,000 galaxies with good estimates of
photometric redshift, and measure the two-point cross-correlation
between quasars and galaxies.  We utilize a novel method developed by
\citet[][hereafter M09]{myer09clust} to derive the projected
real-space projected cross-correlation function, making use of the
full probability distributions for the photometric redshifts.

Throughout this paper we assume a cosmology with $\Omega_{\rm m}=0.3$
and $\Omega_{\Lambda}=0.7$.  For direct comparison with other works,
we assume $H_0=70$ km s$^{-1}$ Mpc$^{-1}$ (except for comoving
distances and dark matter halo masses, which are explicitly given in
terms of $h=H_0/(100$ km s$^{-1}$ Mpc$^{-1})$).  In order to easily
compare to estimated halo masses in other recent works on quasar
clustering \citep[e.g.,][]{croo05, myer06clust, daan08clust,
  ross09qsoclust}, we assume a normalization for the matter power
spectrum of $\sigma_8 = 0.84$.  Photometry is presented in Vega
magnitudes.  All quoted uncertainties are $1\sigma$ (68\% confidence).

\begin{figure*}[t]
\epsscale{1.}
\plottwo{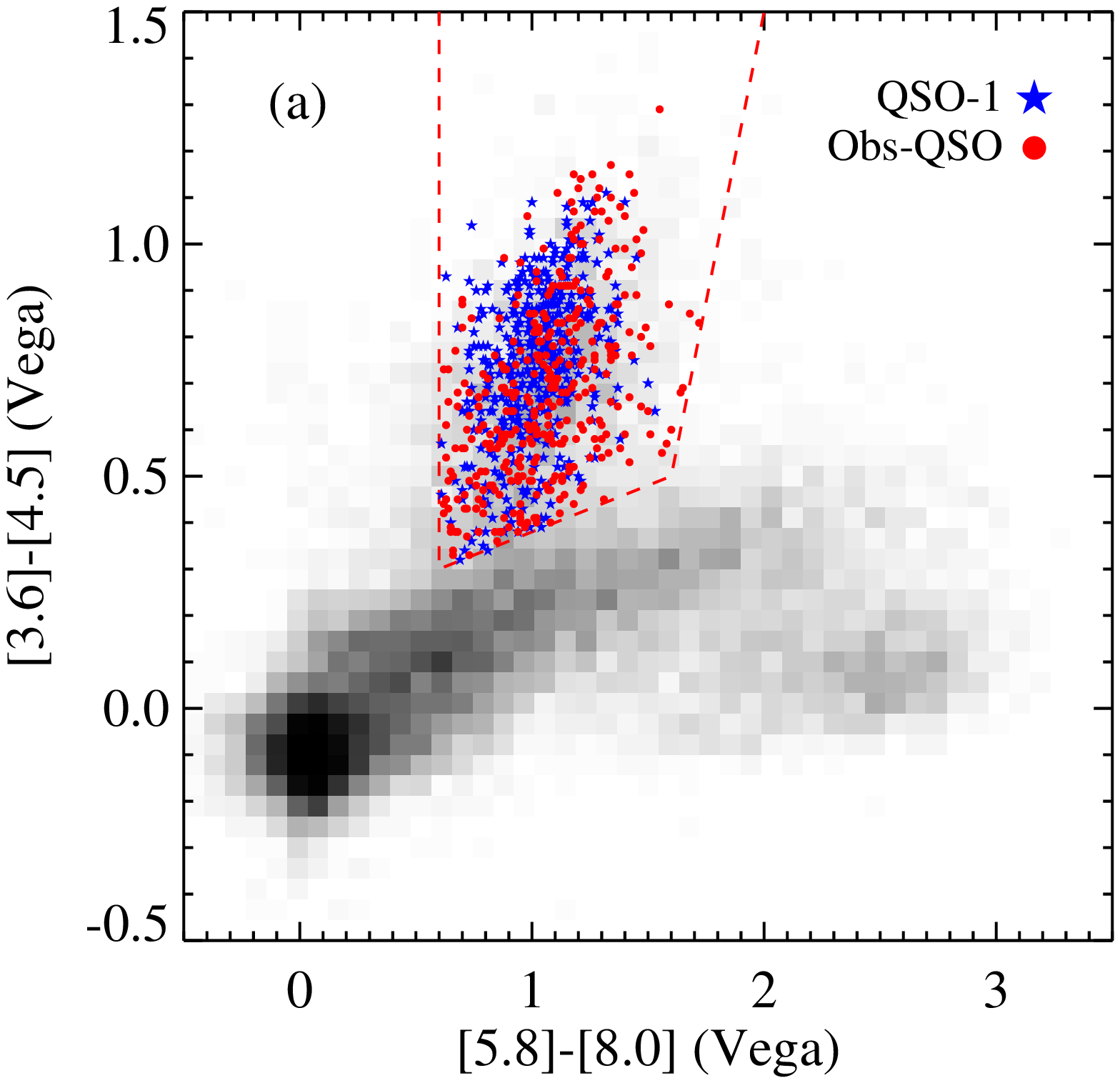}{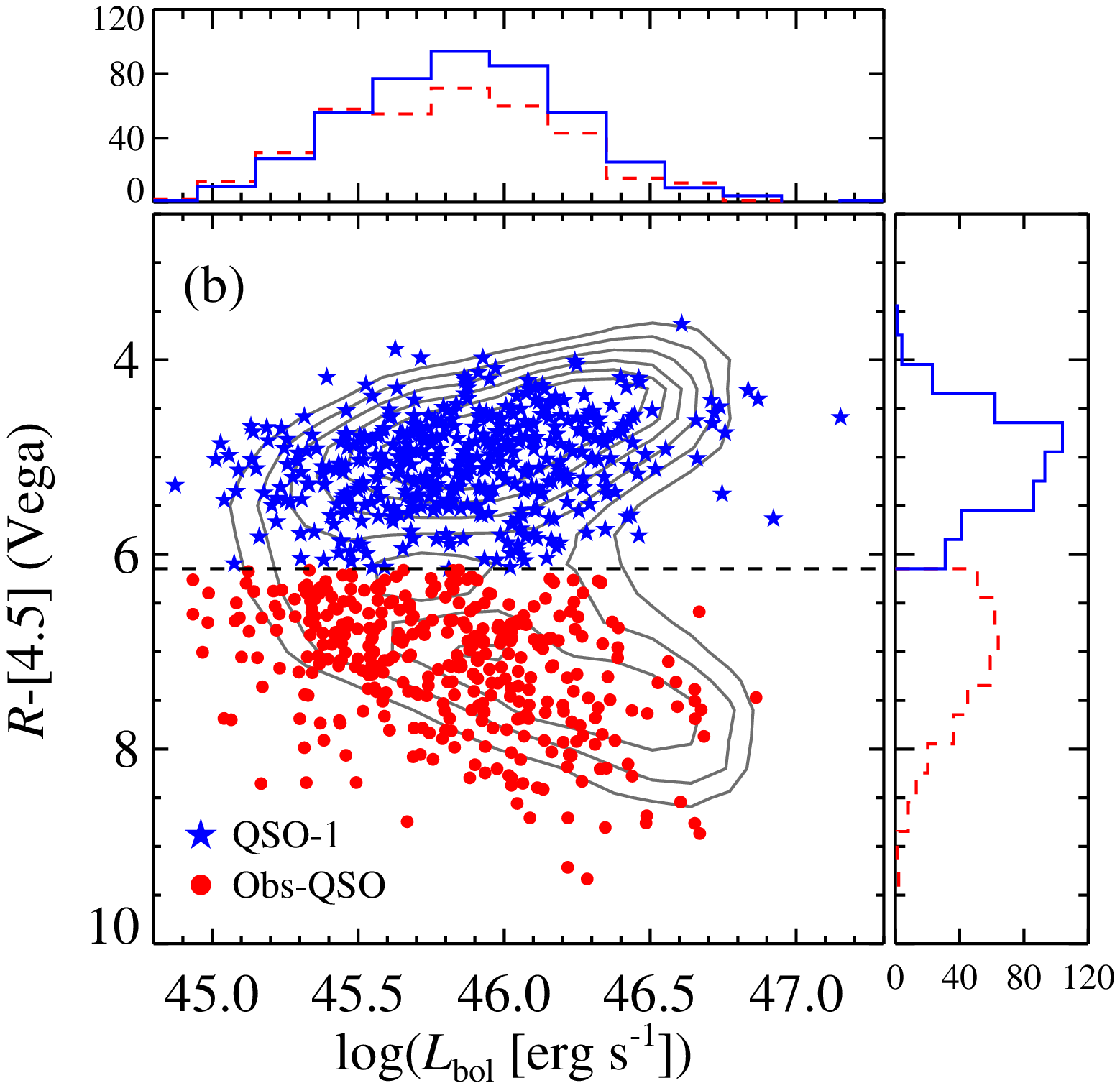}
\caption{(a) IRAC color-color diagram showing the selection of the
  quasar samples using the criteria of \citet{ster05}.  The grayscale
  shows the density of sources detected at $>5\sigma$ significance in
  all four bands in IRAC Shallow Survey data.  Blue stars and red
  circles show the {\qsoone} and {\qsotwo} samples, respectively.  The
  \citet{ster05} color-color selection region is shown by the dashed
  line.    (b) Illustration of the optical-IR
  color-selection criteria for dividing the IR-selected QSO sample
  into unobscured ({\qsoone}) and obscured ({\qsotwo}) subsamples.
  Shown is observed $R-[4.5]$ color versus bolometric luminosity,
  calculated as described in \S~\ref{quasarsample}.  Contours show the
  distribution for all the \citetalias{hick07abs} IR-selected quasars,
  while blue stars and red circles  show the {\qsoone} and {\qsotwo}
  subsamples at $0.7 < z < 1.8$ used in this analysis as described in
  \S~\ref{quasarsample}.  The right and top panels show histograms of color and $L_{\rm bol}$, respectively, for  the {\qsoone}s (blue solid line) and {\qsotwo}s (red dashed line).  The contours and color histograms
  show that a simple cut in optical-IR color clearly separates the
  QSO samples into two populations, while the $L_{\rm bol}$ histograms demonstrate
 that the two samples are very closely matched in luminosity.
 \label{fstern}}
\vspace{1.5ex}
\end{figure*}

\section{Observations}
\label{obs}

The 9 deg$^{2}$ survey region in \bootes\ covered by the NOAO Deep
Wide-Field Survey \citep[NDWFS;][]{jann99} is unique among
extragalactic multiwavelength surveys in its wide field and uniform
coverage using space- and ground-based observatories.  Extensive
optical spectroscopy makes this field especially well suited for
studying the statistical properties of a large number of AGNs
(C.\ Kochanek et~al. 2011, in preparation).  Further details of the \bootes\ data
set have been presented in previous papers \citep[e.g.,][]{hick07abs,
  hick09corr, ashb09sdwfs}.

Redshifts for this study come from the AGN and Galaxy Evolution Survey
(AGES; Kochanek et~al. 2011, in preparation) which used the Hectospec
multifiber spectrograph on the MMT \citep{fabr05}.  We use AGES Data
Release 2 (DR 2), which includes all the AGES spectra taken in
2004--2006.  Details of the AGN redshifts are given in \citetalias{hick07abs}
and \citet{hick09corr}.

Optical photometry from NDWFS was used for the selection of AGES
targets and to derive optical colors and fluxes for AGES sources.
NDWFS images were obtained with the Mosaic-1 camera on the 4-m Mayall
Telescope at Kitt Peak National Observatory, with 50\% completeness
limits of 26.7, 25.0, and 24.9 mag, in the $B_{W}$, $R$, and $I$
bands, respectively.  Photometry is derived using SExtractor
\citep{bert96}.

\begin{figure*}
\epsscale{1.}
\plottwo{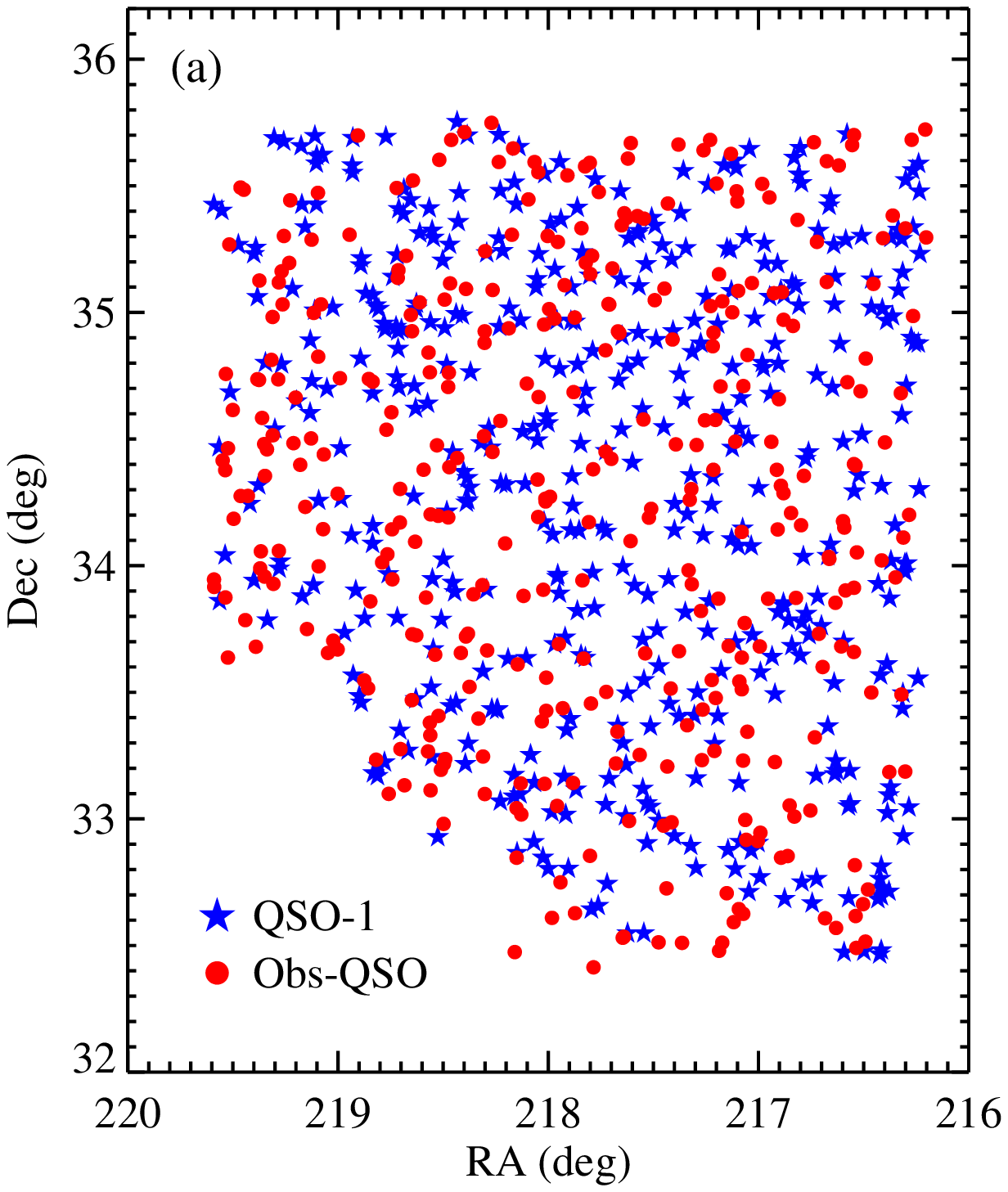}{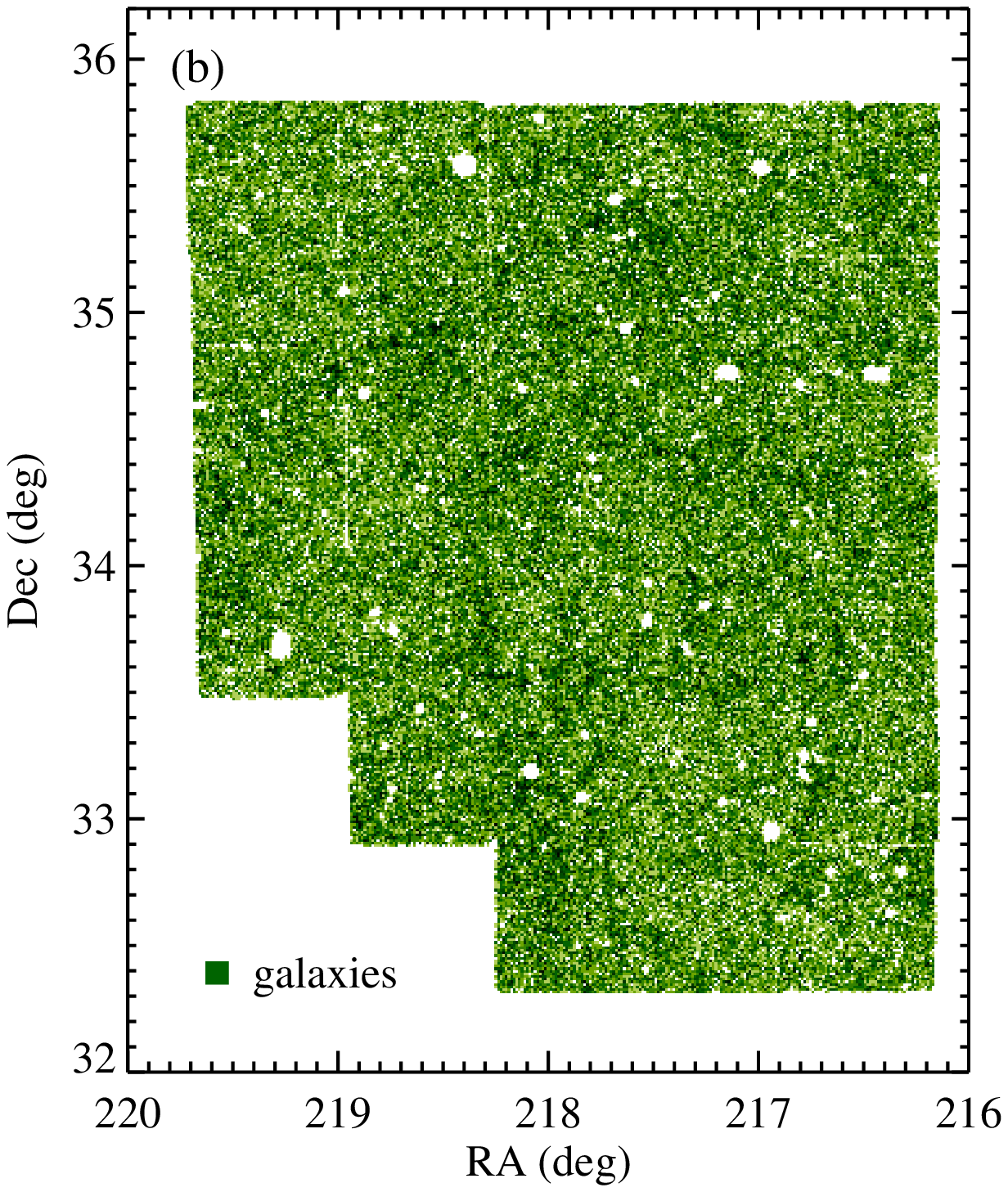}
\caption{{\em Left:} Sky positions of IR-selected quasars at $0. 7 <z
  < 1.8$ in the \bootes\ field.  Quasars are selected using the
  color-color criteria of \citet{ster05}, and are divided into
  unobscured ({\qsoone}; blue stars) and obscured ({\qsotwo}; red circles) as
  sources with optical-IR color blueward and redward, respectively, of
  the boundary $R-[4.5]=6.1$. {\em Right:} Surface density on the sky
  for the sample of 256,124 SDWFS galaxies at $0.5 < z < 2$.  Objects
  in regions of unreliable photometry are excluded
  from both the quasar and galaxy data sets.
\label{fagnsky}}
\vspace{1.5ex}
\end{figure*}

Mid-infrared observations are taken from the \spitzer\ IRAC Shallow
Survey \citep[ISS;][]{eise04}, and {\it Spitzer} Deep Wide-Field
Survey \citep[SDWFS;][]{ashb09sdwfs}.  ISS covers the full AGES field
in all four IRAC bands (3.6, 4.5, 5.8, and 8 $\mu$m), with $5\sigma$
flux limits of 6.4, 8.8, 51 and 50 $\mu$Jy respectively. The IRAC
photometry for ISS is described in detail in \citet{brod06}.  The more
recent SDWFS exposures extend these limits to 3.5, 5.3, 30, and 30
$\mu$Jy, respectively.  As discussed below, the quasar sample
\citepalias[as defined in][]{hick07abs} was selected using ISS data,
while the galaxy sample for cross-correlation is selected from the
full SDWFS data set.  In computing bolometric luminosities for the
quasars, we also make use of 24 \micron\ flux measurements available from
the Multiband Imaging Photometer for {\em Spitzer} (MIPS) GTO
observations (IRS GTO team, J.~Houck (PI), and M.~Rieke) of the
\bootes\ field.  Significant fluxes ($> 3\sigma$) were
obtained for 97\% of the quasars in our sample that lie in the region
covered by MIPS.

\section{Quasar and galaxy samples}
\label{samples}

Our primary analysis is the two-point cross-correlation between
mid-IR selected quasars and galaxies.  In this section we give details of
the the quasar (both obscured and unobscured) and galaxy samples.

\subsection{Quasar sample}

\label{quasarsample}

The quasars\fnm\ are taken from the sample of luminous mid-IR selected
AGNs presented by \citetalias{hick07abs}.  Quasars are identified on
the basis of their colors in the mid-IR as observed by {\it Spitzer}
IRAC, using the color-color criterion of \citet{ster05}
(Figure~\ref{fstern}a), and are selected such that their best
estimates of redshift are at $z > 0.7$.  To the relatively shallow
flux limits of the IRAC Shallow Survey, the AGN sample is highly
complete and suffers little contamination from star-forming galaxies
(as discussed in detail in \S~7 of \citetalias{hick07abs}; see also
\citealt{asse10agntemp, asse11qsolfunc}).  \fnt{We note that while
  \citetalias{hick07abs} refers to the sample as ``AGNs'', their
  bolometric luminosities are estimated to be in the range
  $10^{45}$--$10^{47}$ \ergs, corresponding roughly to an X-ray
  luminosity range $5\times10^{43}\lesssim L_X \lesssim
  5\times10^{45}$ \ergs\ \citep{marc04smbh,hopk07qlf}.  Such high
  luminosity AGNs are typically referred to as ``quasars'' in the
  literature, so to avoid confusion with studies of lower-luminosity
  active galaxies, here we refer to our sample as ``quasars''.}

\citetalias{hick07abs} showed that at the ISS flux limits, the
IR-selected quasars show a bimodal distribution in optical to mid-IR
color.  The selection boundary at $R-[4.5]=6.1$ can be interpreted as
dividing quasars into unobscured (optically bright and so ``blue'' in
$R-[4.5]$) and obscured (optically faint and so ``red'' in $R-[4.5]$)
subsets (Figure~\ref{fstern}b).  For the purposes of this study these
objects will be referred to as ``{\qsoone}s'' and ``{\qsotwo}s'',
respectively; the reader is reminded that the selection is based
not on optical spectroscopy but only on optical to mid-IR color.  This
selection yields samples of 839 {\qsoone}s and 640 {\qsotwo}s at $z > 0.7$.

A detailed study of the optical colors, morphologies, and average
X-ray spectra of these objects is given in \citetalias{hick07abs}. To
briefly summarize, \citetalias{hick07abs} found that the {\qsoone}s
have blue optical colors, point-like optical morphologies, and soft
X-ray spectra characteristic of unobscured quasars, while the
{\qsotwo}s had redder optical colors, extended optical morphologies,
hard X-ray spectra and high $L_X$ characteristic of obscured quasars.
The sample does not include all obscured quasars, as sources with very
large extinction may fall below the IR flux limits of the survey or
move out of the \citet{ster05} selection region (as shown in Figure~1
of \citetalias{hick07abs}; see also \citealt{gorj08,
  asse10agntemp}). The typical absorbing column for the {\qsotwo}
sample is estimated to be $N_{\rm H} \sim 10^{22}$--$10^{23}$ \cdens.
We expect the {\qsotwo}s to suffer little contamination from
bright star-forming galaxies.  \citetalias{hick07abs} used an X-ray
stacking analysis and constraints from deeper surveys to estimate the
possible contamination, and concluded that the contamination is at
most $\approx$30\%, and likely significantly smaller ($<$10\%).

\begin{figure}[t*]
\epsscale{1.1}
\plotone{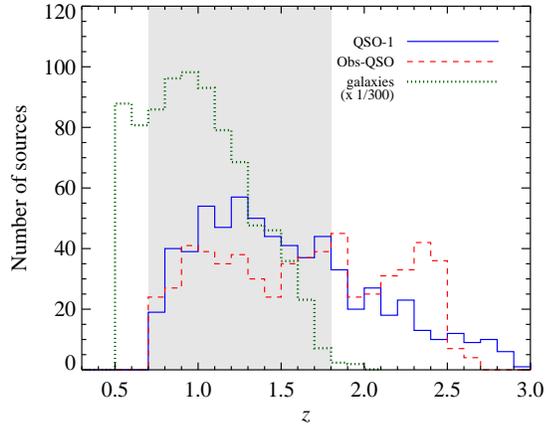}
\caption{Redshift distributions for the photometric galaxy sample
  between at $0.5 < z < 2$ (green dotted line), and the unobscured (blue
  solid line) and obscured (red dashed line) quasar samples of
  \citetalias{hick07abs}.  The histogram for galaxies is multiplied by
  $1/300$ so that the distribution can be directly compared to that of
  the AGNs. The redshift range $0.7 < z < 1.8$ for which the
  correlation analysis is performed is shown by the shaded area.
  Redshift estimates for the galaxies and most of the {\qsotwo}s are
  derived from the photometric redshift calculations using the method
  of \citet{brod06}.  The {\qsoone}s included in the correlation analysis
  have spectroscopic redshifts from MMT/Hectospec observations.
\label{fz}}
\vspace{1.5ex}
\end{figure}

For our spatial correlation analysis, we limit the IR-selected quasar
sample to the redshift range $0.7 < z < 1.8$, to maximize overlap with
the normal galaxies in the field (\S~\ref{galaxy}).  We also include
only objects in regions of good optical photometry and away from
bright stars.  These criteria yield 563 {\qsoone}s and 361 {\qsotwo}s.
Finally, we restrict the {\qsoone} sample to those spectroscopically
identified as broad-line AGNs, to ensure that they unambiguously
represent a sample of unobscured quasars and to enable clean tests of
photo-$z$ errors (see \S~\ref{qsophotoz}).  Of the full sample of
{\qsoone}s all redshifts, the vast majority (80\%) have optical
spectra from AGES and 96\% of these are classified as broad-line AGNs
at  $0.7<z < 4.3$, supporting their selection as unobscured
quasars.  We limit the {\qsoone} sample to the 445 that have accurate
optical spectroscopic redshifts in the range $0.7 < z < 1.8$ and clear
broad emission line features.  (In a sense this is conservative; we
verify that including the 20\% of objects with only photo-$z$s has no
significant effect on the clustering results.)  Based on these
selection criteria, our {\qsoone} sample is essentially equivalent to
other Type 1 quasar samples selected purely on optical photometric
colors and/or spectroscopy \citep[e.g.,][]{rich01qsocol,
  croo04twodfqz, schn07sdssqso, rich09qsophot}, since the vast
majority of spectroscopic Type 1 quasars show AGN-like mid-IR colors
\citep{ster05, rich09bayes}.  The positions on the sky of the final
samples of {\qsoone}s and {\qsotwo}s are shown in
Figure~\ref{fagnsky}(a), and their distribution in redshift is given
in Figure~\ref{fz}.

\begin{figure}[t*]
\epsscale{1.1}
\plotone{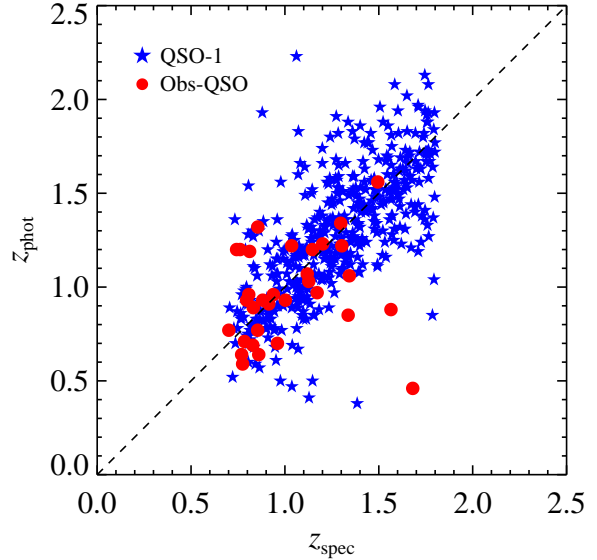}
\caption{Photometric versus spectroscopic redshift for the {\qsoone} (blue
  stars) and {\qsotwo} (red circles) IR-selected quasar samples.  By
  selection, all the {\qsoone}s have spectroscopic redshifts, while the
  {\qsotwo}s were generally too faint for optical spectroscopy.  Only 19
  {\qsotwo}s have good optical redshifts; they are plotted here.  
\label{fspz}}
\vspace{1.5ex}
\end{figure}

The {\qsotwo}s are (by definition) optically faint, and so few (only 7\%)
are bright enough to obtain good redshifts from MMT optical
spectroscopy.  AGES targeted objects down to a flux limit of $I<20$
for sources that are optically extended, which is the case for almost
all the {\qsotwo}s.  Therefore the vast majority of the {\qsotwo} sample has
only photometric estimates of redshift, derived using an artificial
neural net technique \citep{brod06}.  Uncertainties on photo-$z$s
using this technique for optically-bright quasars are typically
$\sigma_{z}=0.12(1+z)$.  However the errors are more difficult to
estimate for optically-faint {\qsotwo}s, for which there are few
spectroscopic redshifts for comparison.  Photo-$z$ uncertainties were
discussed at length by \citetalias{hick07abs}, with the conclusion
that typical uncertainties are at most $\sigma_{z}=0.25(1+z)$ and are
likely smaller.  Figure~\ref{fspz} shows the photo-$z$s and spec-$z$s
for the handful of {\qsotwo}s with spectroscopic redshifts, as well as
those for the {\qsoone}s for comparison.  The impact of photo-$z$ errors
in the present clustering analysis are addressed in detail in
\S~\ref{qsophotoz}. As discussed in \S~\ref{qsophotoz}, random errors
in the photo-$z$s can only tend to {\em decrease} the observed
clustering amplitude, so we expect the present analysis to provide a
robust lower limit on the clustering of the {\qsotwo}s.

Since the primary aim of this analysis is to compare the clustering of
quasars with and without obscuration by dust, it is imperative that
the samples are otherwise matched in key properties such as redshift
and luminosity. We show in Figure~\ref{fz} that the redshift
distributions of the two samples are similar, and we obtain bolometric
luminosities ($L_{\rm bol}$) for the quasars by scaling from the
rest-frame 8 \micron\ luminosity. We compute the flux at rest-frame 8
\micron\ by extrapolating between the fluxes at 8 and 24 \micron\ in
the observed frame, and use this flux to obtain the monochromatic
luminosity $\nu L_\nu$ at 8 \micron.  We then multiply by a
luminosity-dependent bolometric correction from \citet{hopk07qlf},
which ranges from factors of $\approx$8 to 11, in order to obtain
$L_{\rm bol}$. Visual inspection of the {\em Spitzer} data shows that
essentially all of the quasars have broadly power-law SEDs at these
wavelengths, indicating the rest-frame 8 \micron\ emission is indeed
dominated by the AGN.  We note that 49 quasars lie outside the region
covered the MIPS 24 \micron\ observations, while 26 ($\approx$3\%) of
those inside the MIPS area are not detected at 24 \micron. For these
75 objects, we use the estimates of $L_{\rm bol}$ derived from the
rest-frame 2 \micron\ luminosity as in \S~4.6 of
\citetalias{hick07abs}\fnm.

\fnt{In general, the $L_{\rm bol}$ derived from the rest-frame 2
  \micron\ luminosity as used in \citetalias{hick07abs} (which did not
  make use of the 24 \micron\ data) broadly matches that obtain from
  the extrapolated 8 \micron\ flux. However, the median $L_{\rm bol}$
  obtained from 2 \micron\ is smaller for the {\qsotwo}s than for the
  {\qsoone}s by $\approx$0.15 dex, primarily because the {\qsotwo}s
  have somewhat redder mid-IR SEDs consistent with the nuclear
  emission being reddened by dust \citep[e.g.,][]{haas08midirqso}.}

The distributions in $L_{\rm bol}$ are almost identical for the
{\qsoone} and {\qsotwo} samples, as shown in the top panel of
Figure~\ref{fstern}(b). The median and dispersion in $\log{L_{\rm
    bol}}$ (\ergs) is (45.86, 0.37) and (45.83, 0.39) for {\qsoone}s
and {\qsotwo}s, respectively, indicating that the two samples are very
well matched in bolometric luminosity.  For completeness, we note that
if we use the $L_{\rm bol}$ estimates derived from rest-frame 2
\micron\ in \citetalias{hick07abs} and restrict our analysis to
{\qsoone} and {\qsotwo} samples that are matched in $L_{\rm bol}$,
this has a negligible effect on the clustering results.

\subsection{Galaxy sample}

\label{galaxy}

The sample of 256,124 galaxies is selected from the deeper SDWFS IRAC
observations, with a flux limit $[4.5] < 18.6$.  The galaxies are
selected to have best estimates of photometric redshift between 0.5
and 2, with an average photo-$z$ of $\average{z} = 1.09$.  The sample
includes an optical magnitude cut of $I < 24$ to restrict it to
optical fluxes for which the photo-$z$s are well-calibrated.  To
eliminate powerful AGNs, we have also excluded any object detected in
5 ks \chandra\ X-ray observations \citep{kent05} or with 5$\sigma$
SDWFS detections in all four IRAC bands and colors in the
\citet{ster05} AGN selection region.  The exclusion of AGNs from the
galaxy sample removes only 6,979 objects and has negligible effect on
the results.

\begin{figure}
\epsscale{1.1}
\plotone{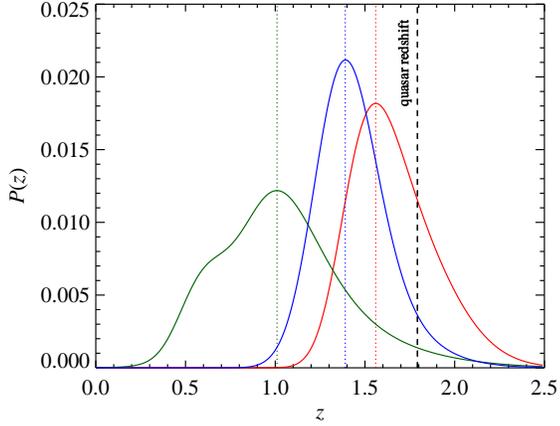}
\caption{Sample probability density functions for three galaxies in
  the SDWFS sample.  Dotted lines show the ``best'' (peak) redshifts
  for each galaxy.  The redshift of a sample quasar is shown by the
  dashed black line.  Note that for the two lower-redshift galaxies,
  the radial distance between the ``peak'' redshift of the galaxy and
  the quasar redshift are far too large for them to be physically
  associated.  However, because of the uncertainty in the galaxy
  redshifts (shown by the PDFs), there is a non-negligible probability
  that the galaxies lie close to the radial distance of the quasar.
\label{fpdf}}
\vspace{1.5ex}
\end{figure}

The distribution on the sky of the 256,124 galaxies are shown in
Figure~\ref{fagnsky}(b), and their distribution in photometric
redshift is shown in Figure~\ref{fz}.  Photometric redshifts are
obtained using an updated version of the \citet{brod06} algorithm,
which is based on template fitting to the optical-IR SEDs.  The SED
fitting produces a redshift probability density function (PDF) for
each object, where $P(z)$ represents the probability that the object
lies at redshift $z$. (Note that the neural net used for the quasar
photo-$z$s does not produce an equivalent estimate of the PDF.  Thus
for the quasars we use the best value for the redshift, as discussed
in \S~\ref{corranal}.)  $P(z)$ is normalized such that $\int P(z) {\rm
  d}z = 1$.  For most galaxies the PDF is roughly Gaussian in shape,
although often with a broader tail toward higher redshift.  The
typical redshift uncertainties are $\sigma_z \sim 0.1 (1+z)$, and only
a small fraction ($\approx$0.6\%) of galaxies show multiple
significant peaks in the PDF at different redshifts.  Typical galaxy
PDFs are shown in Figure~\ref{fpdf}.

In addition to the observed galaxy catalog, the correlation analysis
requires a reference sample of objects with random sky positions, in
order to compare the observed quasar-galaxy pair counts with the
number expected for an uncorrelated distribution.  We use a catalog of
$8\times10^{6}$ random ``galaxies'' that are assigned to random
positions in the regions of good photometry, reflecting the spatial
selection function for the SDWFS galaxies.

\section{Correlation analysis}

\label{corranal}

In this section we outline our methods for measuring the spatial
cross-correlation between quasars and galaxies, the autocorrelation of
the galaxies, and the absolute bias and characteristic dark matter
halo masses.

\subsection{Projected correlation function}

\label{projcorr}

To measure the spatial clustering of quasars, we can in principle
derive the autocorrelation of the quasars themselves, or measure their
{\em cross}-correlation with a sample of other objects (specifically,
normal galaxies) at the same redshifts.  Our quasar sample is too
small to obtain sufficiently good measurements of their
autocorrelation function.  However, cross-correlation with galaxies
(of which there are $\simeq 300$ times as many objects in the
\bootes\ data set) allows far greater statistical power.  Further,
cross-correlation requires knowledge only of the selection function
for the galaxies, which is generally better constrained than that for
AGNs.  Cross-correlations of AGNs with galaxies have proved an
effective technique in a number of previous studies
\citep[e.g.,][]{croo04twodfqz,adel05qsoclust,serb06,li06agnclust,coil07a,
  wake08radio, coil09xclust,hick09corr,padm09qsored, mand09agnclust, moun10qsolrg,
  dono10clust, krum10xclust}.

For the present analysis, the uncertainties in the galaxy photo-$z$s
restrict our ability to perform a full three-dimensional clustering
analysis.  However, making use of the quasar redshifts and the galaxy
photo-$z$ information, we can derive a projected spatial
correlation function ($w_p(R)$, with $R$ in comoving \hmpc) that has
both higher signal-to-noise, and a more straightforward physical
interpretation than, for example, the purely angular correlation
function $\omega (\theta)$.

The two-point correlation function $\xi(r)$ is defined as the
probability above Poisson of finding a galaxy in a volume element $dV$
at a physical separation $r$ from another randomly chosen galaxy, such
that
\begin{equation}
dP=n[1+\xi(r)]dV,
\end{equation}
where $n$ is the mean space density of the galaxies in the sample.
The projected correlation function $w_p(R)$ is defined as the integral
of  $\xi(r)$ along the line
of sight,  
\begin{equation}
\label{eqnwpint}
w_p(R)=2\int_{0}^{\pi_{\rm max}}\xi(R,\pi)d\pi,
\end{equation}
where $R$ and $\pi$ are the projected comoving separations between
galaxies in the directions perpendicular and parallel, respectively,
to the mean line of sight from the observer to the two galaxies.  By
integrating along the line of sight, we eliminate redshift-space
distortions owing to the peculiar motions of galaxies, which distort
the line-of-sight distances measured from redshifts.  $w_p(R)$ has
been used to measure correlations in a number of surveys, for example
SDSS \citep{zeha05a, li06agnclust, myer09clust, krum10xclust}, 2SLAQ
\citep{wake08radio}, DEEP2 \citep{coil07a, coil08galclust,
  coil09xclust}, \bootes\ \citep{hick09corr, star10xclust_aph}, COSMOS
\citep{gill09xclust} and GOODS \citep{gill07c}.

In the range of separations $0.3\lesssim r \lesssim50$ \hminus\ Mpc,
$\xi(r)$ for galaxies and quasars is roughly observed to be a
power-law,
\begin{equation}
\xi(r)=(r/r_0)^{-\gamma}.
\label{eqnplaw}
\end{equation}
  For sufficiently large
$\pi_{\rm max}$ such that we average over all line-of-sight peculiar
velocities, $w_p(R)$ can be directly related to $\xi(r)$ (for a
power law parameterization) by 
\begin{equation}
w_p(R)=R\left (\frac{r_0}{R}\right)^\gamma
\frac{\Gamma(1/2)\Gamma[(\gamma-1)/2]}{\Gamma(\gamma/2)}.
\label{eqnwpplaw}
\end{equation}

We use Equation~(\ref{eqnwpplaw}) to obtain power-law parameters for
the observed correlation functions, to facilitate straightforward
comparisons to other works.  However, we note that a number of recent
studies have shown evidence for separate terms in the correlation
function owing to pairs of galaxies found within a single dark matter
halo (the ``one-halo'' term), and from pairs in which each galaxy is
in a different halo \citep[the ``two-halo'' term; e.g.,][]{zeha04,
  zhen07hod,coil08galclust, brow08halo,zhen09redhod}.  A halo
occupation distribution (HOD) analysis accounting for both the one-
and two-halo terms can provide valuable constraints on the
distribution of objects within their dark matter halos, however a full
HOD calculation is beyond the scope of the present analysis.

\defcitealias{myer09clust}{M09}

To measure $w_p(R)$ for the quasar-galaxy cross-correlation, we employ
the method developed by \citetalias{myer09clust}.  This
technique makes use of the full photo-$z$ PDF for every galaxy, to
weight quasar-galaxy pairs based on the probability of their being
associated in redshift space.  We describe the formalism briefly here,
and refer the reader to \citetalias{myer09clust} for further details.

\subsection{Cross-correlation method}

\label{crosscorr}

For a set of spectroscopic quasars all at the same comoving distance
$\chi_*$ from the observer, the angular cross-correlation between the
(spectroscopic) quasars and (photometric) galaxies can be expressed in
terms of the physical transverse comoving distance by
\citep[e.g.,][]{shan83clust}:
\begin{equation}
w_\theta(R) = \frac{N_R}{N_G}\frac{D_Q D_G (R)}{D_Q R_G (R)} - 1,
\label{eqnccorr}
\end{equation}
where $R$ is the projected comoving distance for a given angular separation $\theta$, such that
$R=\chi_* \theta$.  $N_G$ and $N_R$ are the total numbers of
photometric galaxies and random galaxies, respectively, and $D_Q D_G$
and $D_Q R_G$ are the number of quasar-galaxy and quasar-random pairs
in each bin of $R$.

 Defining the radial distribution function for the full galaxy sample
as $f(\chi)$, where $\int f(\chi) d\chi = 1$, and assuming that
$f(\chi)$ varies slowly at the redshifts of interest, then the angular
correlation function $w_\theta(R)$ is related to the projected real
space correlation function $w_p(R)$ by
\begin{equation}
w_\theta(R) = f(\chi_*)w_p(R)
\end{equation}
 (for a derivation see \S~3.2 of \citealt{padm09qsored}). As discussed in detail in \citetalias{myer09clust}, we can generalize
the analysis such that the contribution to $w_p(R)$ is calculated
individually for each quasar-galaxy pair, with $f_{i,j}$ defined as
the average value of the radial PDF $f({\chi})$ for each photometric object $i$, in a
window of size $\Delta \chi$ around the comoving distance to each
spectroscopic object $j$.  
We use $\Delta \chi = 100$ \hmpc\ to effectively eliminate redshift space distortions,
although the results are insensitive to the details of this choice.  

In this case of weighting by pairs, we obtain, as in Equation (13) of
\citetalias{myer09clust}:
\begin{equation}
w_p(R) = N_R N_Q \sum_{i,j} c_{i,j} \frac{D_Q D_G (R)}{D_Q R_G (R)} - \sum_{i,j} c_{i,j}
\label{eqnwp}
\end{equation}
where
\begin{equation}
\label{eqncij}
c_{i,j} = f_{i,j}\large{/}\sum_{i,j}f_{i,j}^2.
\end{equation}
We refer the reader to \S~2 of \citetalias{myer09clust} for a detailed derivation and discussion of these equations.

\citetalias{myer09clust} use Equation~(\ref{eqnwp}) to compute the
cross-correlation between spectroscopic and photometric quasars from
SDSS in the relatively narrow redshift bin $1.8 < z < 2.2$,
corresponding to comoving distances $3400 \lesssim \chi \lesssim 3800$
\hmpc; they obtain a cross-correlation length $r_0 = 4.56\pm0.48$
\hmpc, assuming $\gamma=1.5$. (Note that \citetalias{myer09clust} do
not derive $\gamma=1.5$, they assume it purely to describe their
method. Higher values of gamma are typically obtained in the recent
literature, and we obtain $\gamma \approx 1.8$ in the present work.)
Our quasar sample spans a comparatively larger range in redshift ($0.7
< z < 1.8$, corresponding to $1750 \lesssim \chi \lesssim 3400$
\hmpc).

We evaluate Equation~(\ref{eqnwp}) by calculating the $D_Q
D_G/D_Q R_G$ term individually for each quasar.  That is, for each
quasar and each bin in separation $R$, we sum the redshift weights
$c_{i,j}$ for galaxies in the given range of distance from the quasar,
and divide by the number of random galaxies in the same distance range
(note that this implies $N_Q = 1$ in Equation~\ref{eqnwp}).  The
advantage of this procedure is that it consists of a simple sum and
accounts exactly for the comoving distance to each quasar.  However,
the calculation is limited by shot noise on small scales where we have
small numbers of quasar-galaxy and quasar-random pairs.  To check that
this does not significantly affect the results, we also divide the
quasar sample into bins of width $\Delta z = 0.1$ (over which the
comoving distance variations are small enough that there is little
mixing between bins in $R$), and calculate the $D_Q D_G/D_Q R_G$ term
for all the $N_Q$ quasars in the bin.  We then average the $w_p(R)$ values for the
different bins to obtain a mean $w_p(R)$ over the redshift range of
interest.  The resulting clustering amplitude differs by
$\lesssim$10\% (and in the majority of cases, $<$ a few percent)
compared to evaluating Equation~(\ref{eqnwp}) treating each quasar
separately. The choice of method does not affect any of our
conclusions, but to account for these differences we conservatively
include an additional 10\% systematic uncertainty on the measurement
of the clustering amplitude. Finally, we emphasize that we are
averaging $w_p(R)$ over the whole redshift range of $0.7 < z < 1.8$.  The
validity of this procedure depends on the fact that the observed
$w_p(R)$ varies slowly in the redshift range of interest, which we
verify explicitly in \S~\ref{variationz}.

\subsection{Galaxy autocorrelation}
\label{galauto}

To estimate dark matter halo masses for the quasars, we calculate the
relative bias between quasars and galaxies from which we derive the
absolute bias of the quasars relative to dark matter.  As discussed
below, calculation of absolute bias (and thus halo mass) requires a
measurement of the autocorrelation function of the SDWFS galaxies.
The large sample size enables us to derive the clustering of the
galaxies accurately from the angular autocorrelation function
$\omega(\theta)$ alone.  Although we expect the photometric redshifts
for the SDWFS galaxies to be well-constrained (as discussed in
\S~\ref{galaxy}), by using the angular correlation function we
minimize any uncertainties relating to individual galaxy photo-$z$s
for this part of the analysis.  The resulting clustering measured for
the galaxies has much smaller uncertainties than that for the
quasar-galaxy cross-correlation.  To save computation time, for the
galaxy autocorrelation analysis we use a significantly smaller random
catalog with only $5\times10^5$ random ``galaxies''.  This likely
introduces some additional shot noise into the calculation of
$\omega(\theta)$, however since the resulting uncertainties are still
far smaller than those for the quasar-galaxy cross-correlation, they
are more than sufficient for the present analysis.

We calculate the angular autocorrelation function $\omega(\theta)$
using the \citet{land93} estimator:
\begin{equation}
\omega(\theta)=\frac{1}{RR}(DD-2DR+RR),
\label{eqnxidef1a}
\end{equation}
where $DD$, $DR$, and $RR$ are the number of data-data, data-random,
and random-random galaxy pairs, respectively, at a separation
$\theta$, where each term is scaled according to the total numbers of
quasars, galaxies, and randoms.

\begin{figure}[t*]
\epsscale{1.1}
\plotone{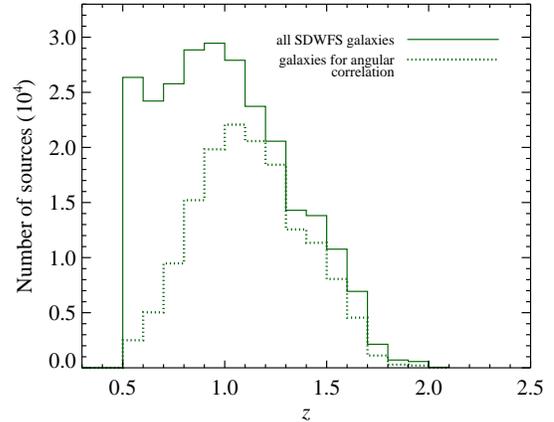}
\caption{Redshift distribution for the full sample of 256,124 SDWFS galaxies  (dotted line) and the sample of 151,256 galaxies selected to match the redshift overlap with the quasars, as described in \S~\ref{galauto} (solid line).  The second galaxy sample is used to derive the angular autocorrelation of the galaxies (\S~\ref{galauto}), as well as their angular cross-correlation with the quasars (\S~\ref{qsoang}).
  \label{fgalselect}}
\vspace{1.5ex}
\end{figure}

The galaxy autocorrelation varies with redshift, owing to the
evolution of large scale structure, and because the use of a
flux-limited sample means we select more luminous galaxies at higher
$z$.  This will affect the measurements of relative bias between
quasars and galaxies, since the redshift distribution of the quasars
peaks at higher $z$ than that for the galaxies and so relatively
higher-$z$ galaxies dominate the cross-correlation signal.  To account
for this in our measurement of galaxy autocorrelation, we randomly
select galaxies based on the overlap of the PDFs with the quasars in
comoving distance (in the formalism of \S~\ref{crosscorr} this is
$f_{i,j}$ for each galaxy, averaged all quasars).  We select the
galaxies so their distribution in redshift is equivalent to the {\em
  weighted} distribution for all galaxies (weighted by
$\average{f_{i,j}}$).  The redshift distribution of this galaxy sample
is shown in Figure~\ref{fgalselect}.  We use this smaller galaxy
sample to calculate the angular autocorrelation of SDWFS galaxies.

\subsection{The integral constraint}

\label{integral}

In fields of finite size, estimators of the correlation function based
on pair counts are subject to the integral constraint, which can be expressed as \citep{grot77}
\begin{equation}
\iint \omega(\theta_{12}) {\rm d}\Omega_1 {\rm d}\Omega_2 \simeq 0,
\end{equation}
 where $\theta_{12}$ is the angle between the solid
angle elements ${\rm d}\Omega_1$ and ${\rm d}\Omega_2$ and the
integrals are over the survey area.  If the number density
fluctuations in the volume are small, and the angular correlations are
smaller than the variance within the volume, then to first order the
correlation function is simply biased low by a constant equal to the
fractional variance of the number counts.  A straightforward way to remove this
bias is to add to the observed $\omega(\theta)$ the term
\begin{equation}
\omega_\Omega = \frac{1}{\Omega^2}\iint \omega(\theta_{12}) {\rm d}\Omega_1 {\rm d}\Omega_2,
\label{eqnint}
\end{equation}
where $\Omega$ is the area of the survey region.  The value of
$\bar{n}^2 \omega_\Omega$, where $\bar{n}$ is an estimate of the mean
number of galaxies per unit area, is the contribution of clustering to
the variance of the galaxy number counts \citep{grot77, efst91clust}.
Evaluating Equation~(\ref{eqnint}) for the \bootes\ survey area and
the typical slope of the $\omega(\theta)$ for the objects considered
here, we obtain $\omega_\Omega \approx 0.03\omega(1\arcmin)$.  We
estimate $\omega(1\arcmin)$ by interpolating the observed
$\omega(\theta)$, then add $\omega_\Omega$ to $\omega(\theta)$ before
performing model fits.  For the projected real-space correlation
functions $w_p(R)$ (which is ultimately derived from individual
estimates of $\omega(\theta)$, as in Equation~\ref{eqnccorr}), we
perform an approximate correction for the integral constraint.  We
determine the value of $w_p(R)$ at the physical scale (typically
0.5--1 \hmpc) corresponding to 1\arcmin\ for each quasar, and add the
average of these estimates (multiplied by 0.03) to the observed $w_p(R)$.
These corrections increase the observed clustering amplitude by
$\approx$10\%, but have little effect on our overall conclusions.

\subsection{Uncertainties and model fits}

\label{uncertainties}

Ideally, uncertainties in $w_p(R)$ and $\omega(\theta)$ would be
determined by calculating the correlation function for various random
realizations of mock IR-selected quasar and galaxy samples, for
example by populating dark matter $N$-body simulations.  In the
absence of such mock catalog, we instead determine uncertainties in
$w_p(R)$ directly from the data through bootstrap resampling.

In a standard bootstrap analysis, the survey volume is divided into
$N_{\rm sub}$ subvolumes, and these subvolumes are drawn randomly
(with replacement) for inclusion in the calculation of the correlation
function.  Owing to the relatively small size of the field compared to
large surveys such as SDSS or 2dF, we are only able to divide the
field into a small number of subvolumes (we choose $N_{\rm sub} = 8$).  The
width of one subvolume corresponds to $\approx$50 \hmpc\ at $z=1.2$,
so that correlations between the subvolumes should be relatively weak.
(We verify explicitly that using a larger $N_{\rm sub}=22$ has no
significant effect on the results.) For each bootstrap sample draw a
total of $3 N_{\rm sub}$ subvolumes (with replacement), which has been
shown to best approximate the intrinsic uncertainties in the
clustering amplitude \citep{norb09clust}.  We then re-calculate
$w_p(R)$ including only the subvolumes in the bootstrap sample.  For
the calculations of $w_p(R)$ we use 10,000 bootstrap samples, for which
the uncertainties at each scale converge to better than 1\%.  (To save
computing time, we limit the analysis to 2000 bootstrap samples for
the angular correlation analyses, for which the uncertainties converge
to within $\approx$1.5\%.)

This bootstrap technique works well for the galaxy autocorrelation,
for which we have a large number of objects and the uncertainty is
dominated by the clustering of the sample rather than counting
statistics.  However, for the quasar-galaxy cross-correlation the
bootstrap analysis results in very small errors that are significantly
smaller than the observed scatter between points.  This appears to be
caused by the fact that, owing to the small quasar samples of only a
few hundred objects, the uncertainties are dominated by shot noise
that is not fully characterized by randomly selecting entire
subvolumes.  To account for the shot noise, we therefore take the sets
of $3 N_{\rm sub}$ bootstrap subvolumes and randomly draw from them
(with replacement) a sample of objects (quasars or galaxies) equal in
size to the parent sample; only pairs including these objects are used
in resulting cross-correlation calculation.  This procedure yields a
good estimate of the shot noise (the resulting $\chi^2_\nu \sim 1$)
while also accounting for covariance due to the large-scale structure.

When fitting power-law models to the observed correlation functions,
we compute parameters by minimizing $\chi^2$, taking into account
covariance between different bins in $R$.  From the bootstrap
analysis, we can estimate the covariance matrix $C_{ij}$ by


\begin{eqnarray}
\label{eqnmij}
 C_{ij} = \frac{1}{1-N} \left[\sum_{k=1}^{N} \bigg(w_p^k(R_i)-w_p(R_i)\bigg)\right. \nonumber \\
          \times\, \bigg(w_p^k(R_j)-w_p(R_j)\bigg)\bigg] \, 
\end{eqnarray}
where $w_p^k(R_i)$ and $w_p^k(R_j)$ are the projected correlation
function derived for the $k$-th bootstrap samples, $N$ is the total
number of bootstrap samples, and $w_p(R)$ is the correlation function
for the full sample.  This formalism is equally valid for bins of
angular separation $\theta$ in calculations of $\omega(\theta)$.  The
1$\sigma$ uncertainty in each bin in $R$ is the square root of the
diagonal component of this matrix ($\sigma_{i}=(C_{ii})^{1/2}$).

Taking into account covariance, $\chi^2$ is defined as
\begin{eqnarray}
 \label{eqnchisq}
 \chi^2 = \sum_{i=1}^{N_{\rm bins}}\sum_{j=1}^{N_{\rm bins}}\bigg(w_p(R_i)-w_p^{\rm model}(R_i)\bigg) \nonumber \\
           \times\,C^{-1}_{ij} \bigg(w_p(R_j)-w_p^{\rm model}(R_j)\bigg)
\end{eqnarray}
where $C^{-1}_{ij}$ is the inverse of the covariance matrix $C_{ij}$.
We determine best-fit parameters by minimizing $\chi^2$, and derive
1$\sigma$ errors in each parameter by the range for which $\Delta
\chi^2 = 1$.  As a check, we also estimate parameter uncertainties by
calculating best-fit parameters for each of the bootstrap samples and
calculating the variance between them; this obtains almost identical
estimates of the errors. Further, we note that if we use only the
diagonal terms in the covariance matrix in determining $\chi^2$, the
variation in the best-fit parameters is significantly smaller than the
statistical uncertainties, indicating that the precise details of the
covariance matrix are relatively unimportant.  

We also note that while in principle the SDWFS field is large enough
to enable measurements of clustering up to $\sim$50 \hmpc\ at $z\sim
1$, we limit the analysis to scales $<12$ \hmpc, because of edge
effects that skew the correlation function on large scales but have
minimal effect on smaller scales.  An investigation of this effect is
given in the Appendix.

\subsection{Power law fits to angular correlation functions}
\label{plawang}

For the projected real-space quasar-galaxy cross-correlation analysis,
we fit power-law models of $w_p(R)$ using Equation~(\ref{eqnwpplaw}).
We also fit power laws to the angular correlation functions (both
galaxy autocorrelations and quasar-galaxy cross-correlations), using
the simple expression
\begin{equation}
\omega(\theta) = A\theta^{-\delta}. 
\label{eqnprojmod}
\end{equation}
For meaningful comparison to other clustering measurements obtained
using samples with different distributions in redshift, we wish to
convert the observed parameters $A$ and $\delta$ to the real-space
$r_0$ and $\gamma$ as defined in Equation~(\ref{eqnplaw}).  Inverting
Limber's equation, the conversion between these parameters can be
computed analytically (here we follow \S~4.2 of \citealt{myer06clust}; for the full derivation see \citealt{peeb80}):

\begin{eqnarray}
\gamma&=& \delta +1 \label{eqndelta}\\
A &=&  H_\gamma \frac{\int^{\infty}_0({\rm d}N_1/{\rm d}z)({\rm d}N_2/{\rm d}z) E_z \chi^{1-\gamma}{\rm d}z}{\left[\int^{\infty}_0({\rm d}N_1/{\rm d}z){\rm d}z\right]\left[\int^{\infty}_0({\rm d}N_2/{\rm d}z){\rm d}z\right]} r_{0}^\gamma
\label{eqndeproj}
\end{eqnarray}
where $H_\gamma=\Gamma(0.5)\Gamma\left(0.5[\gamma-1]\right)/\Gamma(0.5\gamma)$, $\Gamma$ is the gamma function, $\chi$ is the radial comoving distance, ${\rm d}N_{1,2}/{\rm d}z$ are the redshift distributions of the samples (for an autocorrelation ${\rm d}N_1/{\rm d}z = {\rm d}N_2/{\rm d}z$), and $E_z = H_z/c = {\rm d}z/{\rm d}\chi$.  The Hubble parameter $H_z$ can be found via 

\begin{equation}
H_z^2 = H_0^2\left[\Omega_m(1+z)^3+\Omega_\Lambda\right].
\label{eqnhz}
\end{equation}

Equation (\ref{eqndeproj}) assumes no evolution with redshift in the
clustering of the sample (equivalent to the implicit assumption made
in fitting $w_p(R)$ with Equation~\ref{eqnwpplaw}).  For each angular
correlation analysis, we derive $A$ and $\delta$ from the observed
$\omega(\theta)$ and then obtain the corresponding $\gamma$ and $r_0$
from Equations~(\ref{eqndelta}) and (\ref{eqndeproj}).

\begin{figure*}[t]
\epsscale{0.75}
\plotone{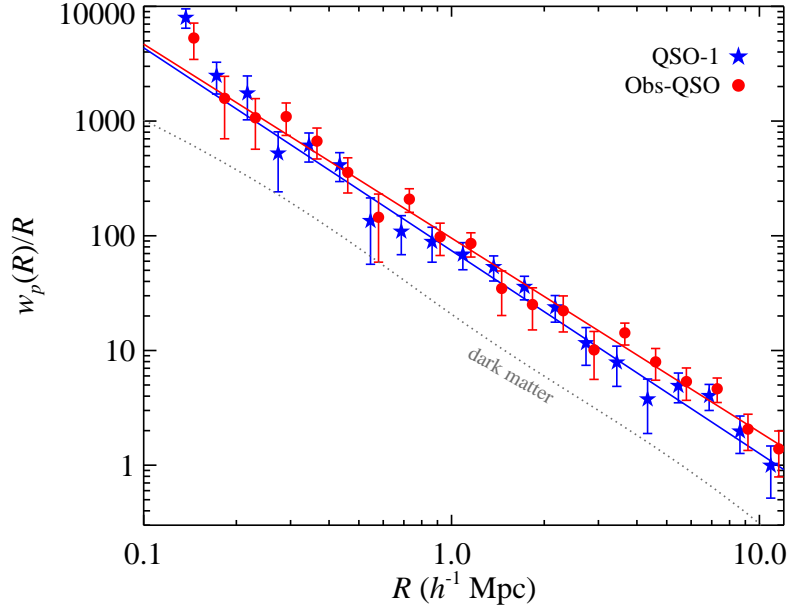}
\caption{The projected quasar-galaxy cross-correlation function
  (derived using Equation~\ref{eqnwp}) for the {\qsoone} (blue stars) and
  {\qsotwo} (red circles) samples.  Uncertainties are estimated
  from bootstrap resampling.  Data points for the two quasar types are
  slightly offset for clarity.  Power-law fits to $w_p(R)$ are shown
  as solid lines in blue ({\qsoone}s) and red ({\qsotwo}s), and the projected
  correlation function for dark matter is shown by the dotted gray
  line.  Fits are performed over the range in separation of $0.3 <
  R < 12$ \hmpc.
\label{fcorr}}
\vspace{1.5ex}
\end{figure*}

\subsection{Absolute bias and dark matter halo mass}
\label{absbias}
The masses of the dark matter halos in which galaxies and quasars
reside are reflected in their absolute clustering bias relative
to the dark matter distribution.  To determine absolute bias
\citep[following e.g.,][]{myer07clust1, coil08galclust, hick09corr} we
first calculate the two-point autocorrelation of dark matter as a
function of redshift.  We use the {\tt HALOFIT} code of
\citet{smit03dm} to determine the nonlinear-dimensionless power
spectrum $\Delta^2_{\rm NL}(k,z)$ of the dark matter assuming our
standard cosmology, and the slope of the initial
fluctuation power spectrum, $\Gamma=\Omega_m h=0.21$.  The Fourier
transform of the $\Delta^2_{\rm NL}(k,z)$ gives us the real-space
correlation function $\xi(r)$, which we then integrate to $\pi=100$
\hminus\ Mpc following Equation (\ref{eqnwpint}) to obtain the dark
matter projected correlation function $w_p^{\rm DM}(R,z)$.  The
uncertainty in the DM power spectrum obtained from {\tt HALOFIT} is
$\sim5$\%; this corresponds to a systematic uncertainty $\sim$0.05 dex
in $M_{\rm halo}$, but does not impact the  relative halo masses
of the different subsamples.

To derive quasar absolute bias from the projected real-space
correlation function, we average the $w_p^{\rm DM}(R)$ over the
redshift distribution of the sample, weighted by the overlap with the
galaxy PDFs.  The overlap of each quasar with the galaxy PDFs is given
by

\begin{equation}
W_i = \sum_{j} f_{i,j}
\end{equation}
and the corresponding $w_p(R)$ for the dark matter is given by
\begin{equation}
w_p^{\rm DM}(R) = \sum_i W_i w_p^{\rm DM}(R,z_i) / \sum_i W_i
\end{equation}
where $z_i$ is the quasar redshift.

The redshift distributions for the {\qsoone}s and {\qsotwo}s are essentially
identical (the resulting $w_p^{\rm DM} (R)$ values for the two samples
differ by $<$2\% on all scales) so for simplicity we use the same
$w_p^{\rm DM}(R)$ (defined for the {\qsoone}s) for both sets of quasars.
We obtain the bias by calculating the average ratio between the
best-fit power-law model and $w_p^{\rm DM}(R)$ over the range of
scales of 1--10 \hmpc, for which $w_p^{\rm DM}(R)$ corresponds closely
to a power law and is dominated by the two-halo term. The observed
clustering amplitude relative to the dark matter corresponds to $b_Q
b_G$, where $b_Q$ and $b_G$ are the absolute linear biases of the
quasars and SDWFS galaxies, respectively.

To measure $b_G$ from the galaxy autocorrelation function (or $b_Q
b_G$ from the quasar-galaxy angular cross-correlation, described in
\S~\ref{qsoang}), we require an estimate of the corresponding
$\omega(\theta)$ of the dark matter.  To obtain $\omega(\theta)$ we
use Limber's equation to project the power spectrum $\Delta^2_{\rm
  NL}(k,z)$ into the angular correlation
\citep{limb53,peeb80,peac91clust,baug93clust}.  Specifically, we
perform a Monte Carlo integration of Equation (A6) of
\citet{myer07clust1} to obtain $\omega(\theta)$ for the dark matter.
The key parameter in this equation is $({\rm d}N_G/{\rm d}z)^2$ where
${\rm d}N_G/{\rm d}z$ is the redshift distribution of the galaxies.
We calculate ${\rm d}N_G/{\rm d}z$ from the sum of the PDFs of the
galaxies for which we perform the autocorrelation.  In deriving the
dark matter $\omega(\theta)$ for the quasar-galaxy cross-correlation,
we replace $({\rm d}N/{\rm d}z)^2$ with $({\rm d}N_Q/{\rm d}z)({\rm
  d}N_G/{\rm d}z)$ where ${\rm d}N_Q/{\rm d}z$ is the distribution of
quasar redshifts.  For each angular correlation analysis we compute
the average ratio between the best-fit power law model and the dark
matter $\omega(\theta)$ on scales 1\arcmin--10\arcmin, where
$\omega(\theta)$ is dominated by the two-halo term.  This ratio
yields $b_G^2$ for galaxy autocorrelations or $b_Q b_G$ for
quasar-galaxy cross-correlations.

Finally, we use $b_Q$ and $b_G$ to estimate the characteristic
mass of the dark matter halos hosting each subset of galaxies or
quasars.  \citet{shet01halo} derive a relation between dark-matter
halo mass and large-scale bias that agrees well with the results of
cosmological simulations.  We use Eqn.~(8) of \citet{shet01halo} to
convert $b_{\rm abs}$ to $M_{\rm halo}$ for the mean redshift of each
subset of objects.  If we use a different relation between $b_{\rm
  abs}$ and $M_{\rm halo}$ \citep{tink05}, we obtain estimates for
$M_{\rm halo}$ that are similar, although slightly larger by 0.2--0.3
dex; these differences do not significantly affect our conclusions.

We note that to estimate $M_{\rm halo}$, we have performed fits to the
observed $w_p(R)$ on scales of 0.3--12 \hminus\ Mpc.  In principle the
dark matter and galaxy correlation functions can have somewhat
different shapes such that the bias depends on the range of scales
considered.  If we limit the fits on scales 1--12 \hmpc, the results
change by $\lesssim$5\%, but with slightly larger uncertainties.  We
also note that our estimates of $M_{\rm halo}$ are relatively
insensitive to our choice of $\sigma_8$.  If we change $\sigma_8$ from
0.84 to 0.8 \citep[as favored by the more recent WMAP cosmology,
  e.g.][]{sper07wmap3} our $M_{\rm halo}$ estimates for quasars and
galaxies increase by $\approx0.1$ dex.

\section{Results}
\label{results}

In this section we discuss the results of the correlation analysis and
the characteristic dark matter halo masses for galaxies and quasars.
We first calculate the cross-correlation of the full {\qsoone} and {\qsotwo}
samples with SDWFS galaxies.  The resulting $w_p(R)$ values and
best-fit models are shown in Figure~\ref{fcorr}, and fit parameters
are given in Table~\ref{tabcorr}.  For both sets of the quasars the
observed real-space projected cross-correlation is highly significant
on all scales from 0.1--12 \hmpc, and the power law fits return
$\gamma\approx 1.8$, similar to many previous correlation function
measurements for quasars \citep[e.g.,][]{coil07a, ross09qsoclust} and
galaxies \citep[e.g.,][]{zeha05a, coil08galclust}.  The best-fit
parameters are $r_0=5.4\pm0.7$ \hmpc, $\gamma=1.8\pm0.1$ for the {\qsoone}s, and $r_0 = 6.4\pm0.8$ \hmpc, $\gamma=1.7\pm0.1$ for the {\qsotwo}s.
The results indicate that the cross-correlation of the {\qsotwo}s with
galaxies is somewhat stronger than that for the {\qsoone}s.  The corresponding values of $b_Q b_G$ are given in Table~\ref{tabcorr}.

As a check,
we also perform power law fits to $w_p(R)$ but leaving the slope fixed
to $\gamma=1.8$, which corresponds to the slope of the $w_p(R)$ for
the dark matter.  This also yields a significant difference in the
clustering amplitude, although somewhat smaller, with $r_0 =
5.3\pm0.6$ and $r_0 = 6.0\pm0.6$ \hmpc\ for the {\qsoone}s and {\qsotwo}s,
respectively.  (Note that the formal uncertainties in $r_0$ here are
smaller than for the above results because they do not account for
covariance with $\gamma$.)

\begin{figure}[t]
\epsscale{1.1}
\plotone{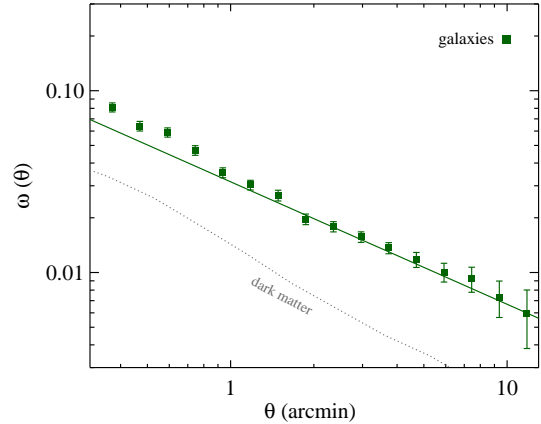}
\caption{ The angular autocorrelation function of SDWFS galaxies,
  selected to match the overlap of the quasars and galaxies in
  redshift space.  Uncertainties are estimated from bootstrap resampling.
  The angular correlation function for dark matter, evaluated for the
  redshift distributions of the galaxies, is shown by the dotted gray
  line.  The power law fit was performed on scales 1\arcmin--12\arcmin\ and is shown as the solid line.  The excess in $\omega(\theta)$ at $\theta \lesssim 1$\arcmin\ is due to the one-halo term arising from pairs of galaxies within the same dark matter halos \citep[e.g.,][]{quad08eroclust, kim11eroclust}.
  \label{fgalang}}
\vspace{1.5ex}
\end{figure}

To obtain the absolute bias of SDWFS galaxies ($b_G$) in order to
extract the quasar bias $b_Q$ from the cross-correlation results, we
next derive the autocorrelation of SDWFS galaxies for the sample
described in \S~\ref{galauto}.  The observed $\omega(\theta)$ is shown
in Figure~\ref{fgalang}, along with the correlation function for dark
matter, calculated as discussed in \S~\ref{absbias}. Fit
parameters are given in Table~\ref{tabcorr}. The power-law model fits
well on the chosen scales of 1\arcmin--12\arcmin, although there is a
clear excess corresponding to the one-halo term at $\theta <
1$\arcmin, as is common in galaxy autocorrelation measurements
\citep[e.g.,][]{quad08eroclust, kim11eroclust}.  The best-fit power
law parameters are $r_0 = 4.7 \pm 0.2$ and $\gamma = 1.67\pm0.05$, and the
ratio of the best-fit power law to the dark matter $\omega(\theta)$
yields $b_G^2 = 2.79\pm 0.16$ or $b_G = 1.67\pm 0.05$.

This accurate value for $b_G$ allows us to estimate $b_Q$ for both
types of quasars, based on the cross-correlation measurements.  We obtain $b_Q =
2.17 \pm 0.55$ and $3.06 \pm 0.70$, for {\qsoone}s and {\qsotwo}s, respectively.
Converting these to dark matter halo masses using the prescription of
\citet{shet01halo} as described in \S~\ref{absbias}, we arrive at
$\log{(M_{\rm halo} [h\; M_{\sun}^{-1}])}= 12.7^{+0.4}_{-0.6}$ and
$13.3^{+0.3}_{-0.4}$ for {\qsoone}s and {\qsotwo}s, respectively.  The difference
is marginally significant ($\approx 1 \sigma$, although as we
discuss below, the {\qsotwo} clustering may represent only a robust lower
limit).

For direct comparison with other studies that directly measure the
quasar autocorrelation, it is useful to present the quasar clustering
in terms of effective power law parameters for their autocorrelation.
Assuming linear bias, the quasar autocorrelation can be inferred from
the cross-correlation by $\xi_{QQ} = \xi_{QG}^2 / \xi_{GG}$
\citep[e.g.,][]{coil09xclust}.  We can therefore use the power law
fits to the quasar-galaxy cross-correlation and galaxy autocorrelation
to derive an effective $r_0$ and $\gamma$ for the quasar
autocorrelation.  This yields $r_0 = 6.1^{+1.4}_{-1.6}$ \hmpc\ and
$\gamma = 1.9^{+0.3}_{-0.2}$ for the {\qsoone}s and $r_0 =
8.8^{+2.0}_{-2.3}$ \hmpc\ and $\gamma = 1.7\pm0.2$ for the {\qsotwo}s.
The autocorrelation amplitude and $M_{\rm halo}$ for {\qsoone}s are in
excellent agreement with previous estimates for unobscured quasars,
while the best-fit amplitude for {\qsotwo}s is higher than most previous
measurements of quasar clustering.  We compare these results to
previous work and discuss possible interpretations in
\S~\ref{discussion}.

\begin{deluxetable*}{lcccccccc}
\tabletypesize{\footnotesize}
\tablewidth{7in}
\tablecaption{Correlation results \label{tabcorr}}
\tablehead{
\colhead{} &
\colhead{} &
\colhead{} &
\multicolumn{3}{c}{Power law fit} &
\multicolumn{2}{c}{Bias\tnm{c}} &
\colhead{Halo mass\tnm{c}}  \\
\colhead{Subset} &
\colhead{$N_{\rm src}$\tnm{a}} &
\colhead{$\average{z}$\tnm{b}} &
\colhead{$r_0$ ($h^{-1}$ Mpc)} &
\colhead{$\gamma$} &
\colhead{$\chi^2_\nu$} &
\colhead{$b_{\rm abs}b_G$} &
\colhead{$b_{\rm abs}$} &
\colhead{($\log{h^{-1} M_{\sun}}$)}}
\startdata
\multicolumn{9}{c}{\em Projected real-space cross-correlation ($w_p(R)$)\tnm{d}} \\
{\qsoone} & 445 & 1.27 & $5.4\pm0.7$ & $1.8\pm0.1$ & 1.1 & $3.63\pm0.92$ & $2.17\pm0.55$ & $12.7^{+ 0.4}_{- 0.6}$ \\
{\qsotwo} & 361 & 1.24 & $6.4^{+0.7}_{-0.8}$ & $1.7\pm0.1$ & 1.2 & $5.11\pm1.16$ & $3.06\pm0.70$ & $13.3^{+ 0.3}_{- 0.4}$ \\
\multicolumn{9}{c}{\em  $w_p(R)$ with fixed $\gamma$\tnm{d}} \\
{\qsoone} & 445 & 1.27 & $5.3\pm0.6$ & 1.8 & 1.1 & $3.44\pm0.75$ & $2.06\pm0.45$ & $12.6^{+ 0.4}_{- 0.5}$ \\
{\qsoone} (photo-$z$) & 445 & 1.27 & $4.9\pm0.6$ & 1.8 & 1.2 & $3.05\pm0.67$ & $1.82\pm0.40$ & $12.5^{+ 0.4}_{- 0.6}$ \\
{\qsotwo} & 361 & 1.24 & $6.0\pm0.7$ & 1.8 & 1.2 & $4.38\pm0.84$ & $2.62\pm0.51$ & $13.0^{+ 0.3}_{- 0.4}$ \\
\multicolumn{9}{c}{\em Angular correlation ($\omega(\theta)$)\tnm{e}} \\
galaxies & 151256 & 1.10 & $4.7\pm0.2$ & $1.67\pm0.05$ & 1.1 & $2.79\pm0.16$ & $1.67\pm0.05$ & $12.30\pm 0.06$ \\
{\qsoone} & 445 & 1.27 & $5.6\pm0.8$ & 1.8 &1.1 & $4.19\pm1.07$ & $2.50\pm0.65$ & $13.0^{+ 0.4}_{- 0.6}$ \\
{\qsotwo} & 361 & 1.24 & $6.0\pm1.0$ & 1.8 &1.2 & $4.80\pm1.29$ & $2.87\pm0.77$ & $13.2^{+ 0.3}_{- 0.5}$
\enddata

\tnt{a}{Number of objects include in the correlation analysis.  For quasar-galaxy cross-correlation, we use the full
  sample of 256,124 galaxies (for $w_p(R)$ calculations) or 151,256
  galaxies (for $\omega(\theta)$ calculations).}
\tnt{b}{Median redshift  for the objects included in the correlation
  analysis. }

\tnt{c}{Uncertainties in the DM power spectrum introduce an additional
  systematic error of $\sim$5\% in $b_{\rm abs}$ (and corresponding
  $\sim$0.05 dex in $M_{\rm halo}$).  Further systematic errors in
  $M_{\rm halo}$ of $\sim$0.2 dex are caused by uncertainty in
  $\sigma_8$ and in the conversion from $b_{\rm abs}$ to $M_{\rm
    halo}$, as discussed in \S~\ref{absbias}.  However, these do not
  significantly effect the {\em relative} halo masses, so these
  uncertainties is not included here.  Note that for fits with fixed
  $\gamma$, uncertainties on $r_0$, bias, and $M_{\rm halo}$ do not
  account for covariance with $\gamma$ and thus somewhat underestimate
  the error on the clustering amplitude.}

\tnt{d}{Real-space projected cross-correlation between quasars and
  galaxies, calculated as described in \S~\ref{corranal}. For all
  $w_p(R)$ calculations, error estimates for $r_0$, bias, and $M_{\rm
    halo}$ include a 10\% systematic uncertainty on the amplitude as described in \S~\ref{crosscorr}.}

\tnt{e}{Angular galaxy autocorrelation and quasar-galaxy
  cross-correlation, calculated as described in \S~\ref{corranal}.}

\vspace{1.5ex}
\end{deluxetable*}

\section{Verification}

\label{verification}

In this section we perform several tests to verify the validity of the
clustering analysis outlined in \S~\ref{corranal}.  We first calculate
the quasar-galaxy cross-correlation using a simple angular correlation
function, minimizing dependence on the photometric redshifts.  We
then check for variation in the observed clustering over the redshift
range of interest and confirm that any variation is relatively weak.
Finally, we estimate the effects of uncertainties on the photometric
redshifts on the observed real-space clustering amplitude for the
{\qsotwo}s.  These checks confirm that our projected correlation
analysis provides a robust estimate of the quasar-galaxy
cross-correlation.

\subsection{Angular cross-correlation}

\label{qsoang}

We first calculate the cross-correlation of quasars and SDWFS galaxies
using a simple angular clustering analysis, and check whether the
corresponding absolute bias is consistent with that derived from the
more sophisticated $w_p(R)$ calculation.  To calculate the
$\omega(\theta)$ we use an estimator corresponding to
Equation~(\ref{eqnxidef1a}) but for cross-correlations:
\begin{equation}
\omega(\theta)=\frac{1}{RR}(D_QD_G-D_QR-D_GR+RR),
\label{eqnxidef2}
\end{equation}
where each term is scaled according to the total numbers of
galaxies and randoms. To maximize the signal-to-noise ratio by
cross-correlating objects associated in redshift space, the galaxies
include only the redshift-matched SDWFS sample of 151,256 objects
described in \ref{galauto}.  Uncertainties are estimated using
bootstrap resampling as described in \S~\ref{uncertainties}.  We fit
the observed cross-correlation with a a power law as described in
\S~\ref{plawang}. Owing to the limited statistics which provide only
very weak constraints on the power law slope, we fix $\delta = 0.8$
(corresponding to real-space $\gamma = 1.8$).

\begin{figure}
\epsscale{1.1}
\plotone{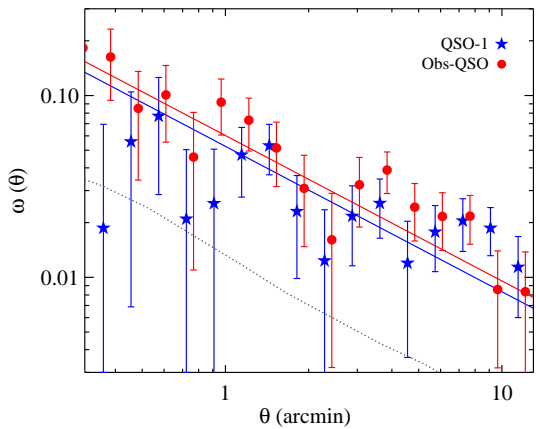}
\caption{The angular quasar-galaxy cross-correlation function for the
  {\qsoone} (blue stars) and {\qsotwo} (red circles) samples.
  Uncertainties are estimated from bootstrap resampling.  Data points for the
  two quasar types are slightly offset for clarity. The angular
  correlation function for dark matter, evaluated for the redshift
  distributions of the galaxies and the {\qsoone}s, is shown by the dotted gray line,  and power law fits (with fixed $\delta=0.8$) are shown as solid blue
  and red lines.  All fits are performed over the range in separation
  of 1\arcmin\ to 12\arcmin.
\label{fqsoang}}
\vspace{1.5ex}
\end{figure}

The resulting cross-correlations and scaled dark matter fits are shown
in Figure~\ref{fqsoang}, and fit parameters are given in
Table~\ref{tabcorr}.  The estimates of $r_0$ and $b_Q b_G$ are in
broad agreement between the two estimators, although as may be
expected, the statistical uncertainties for the angular correlation
analysis are larger (by $\sim$50\%) than for the real-space analysis with fixed
$\gamma$.  Given that the absolute bias derived from the
projected correlation function corresponds broadly to the bias from
the noisier, but simpler angular cross-correlation, we conclude that
there are no significant systematic effects that skew our estimate of
$w_p(R)$.

\subsection{Variation in $w_p(R)$ with redshift}
\label{variationz}

Our calculation of the real-space quasar-galaxy correlation function
over the redshift range $0.7 < z < 1.8$ requires that $w_p(R)$ varies
only slowly between these redshifts, as discussed in
\S~\ref{crosscorr}.  If the objects reside in similar halos at all
redshifts, then we may expect $r_0$ to change slowly; simulations
suggest that the typical $r_0$ for the autocorrelation of DM halos of
mass $\sim$$10^{12}$--$10^{13}$ \hmsun\ should change by $\lesssim
0.2$ \hmpc\ between $z=0.5$ and 2 \citep[see Figure 10
  of][]{star10xclust_aph}.  To test explicitly the redshift variation for the
clustering in our sample, we re-derived $w_p(R)$ using the method
outlined in \S~\ref{corranal} but selecting quasars over smaller
redshift bins of $0.7 < z < 1.25$ and $1.25 < z < 1.8$.  Uncertainties
are calculated using the bootstrap method as for the full quasar
samples, and dark matter and power-law fits are again performed over
the range of separations $0.3 < R < 12$ \hmpc.  We evaluate
$w_p^{DM}(R)$ as in \S~\ref{absbias}, but only including the quasars
in the redshift ranges of interest.  Owing to larger statistical
errors and for simple comparison to the results over the full redshift
range, in the power law fits we fix $\gamma$ to 1.8.

\begin{figure*}
\epsscale{1.1}
\plottwo{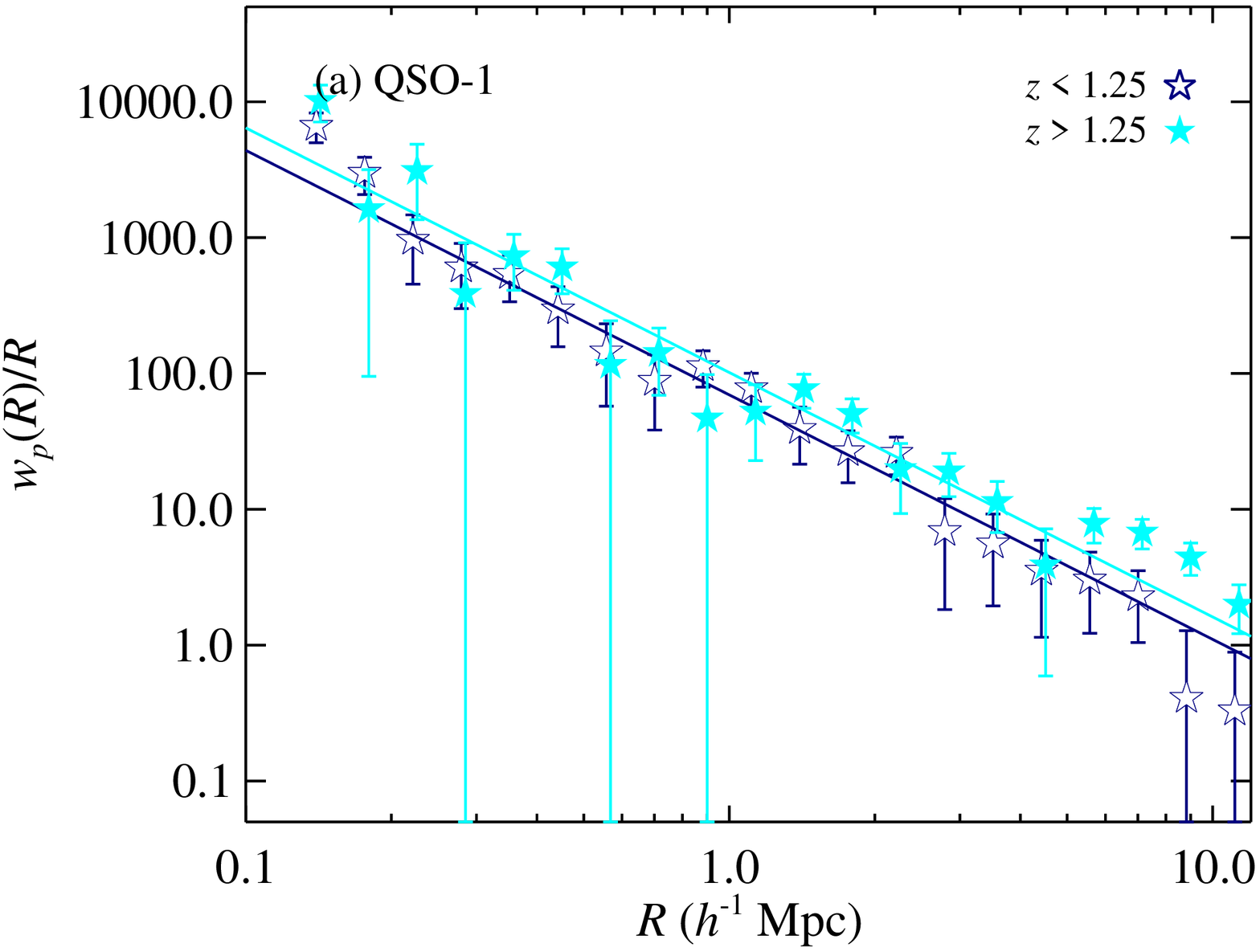}{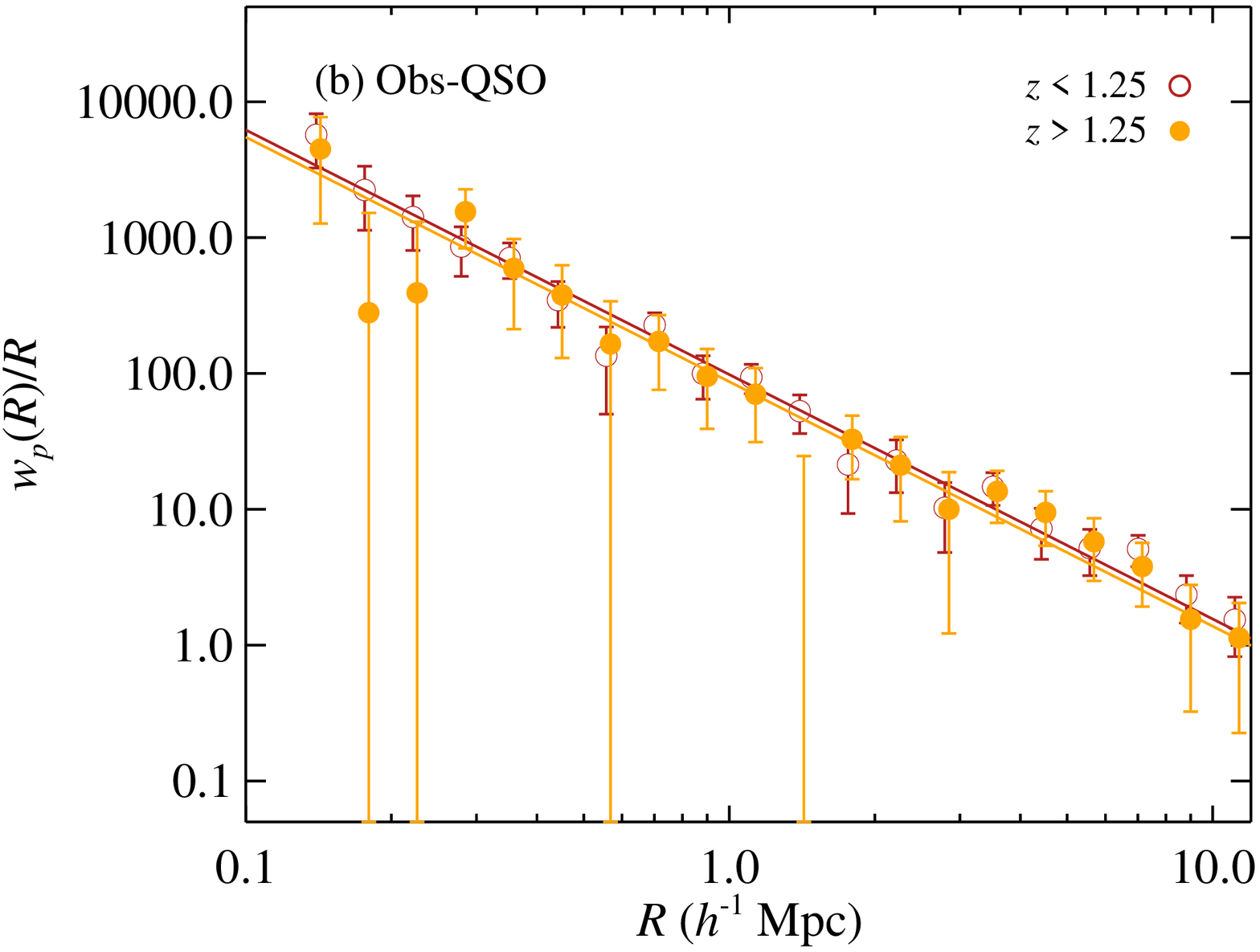}
\caption{Projected quasar-galaxy cross-correlation function in two
  redshift bins: $0.7 < z < 1.25$ and $1.25 < z < 1.8$, with symbols
  shown as in the upper right.  Data points for the two redshift bins
  are slightly offset for clarity.  Power-law fits with $\gamma$ fixed to 1.8
  are shown as solid lines, with colors corresponding to the symbols
  for each redshift bin.  Any variation in $w_p(R)$ with redshift is
  relatively weak, confirming that the method outlined in
  \S~\ref{crosscorr} should be valid for this analysis.  Note that for the {\qsotwo}s at $z>1.25$, we exclude the bin with $w_p(R)<0$, which disproportionately affects the fit. (Including this bin decreases the clustering amplitude by $\approx$20\%.)  Fit parameters are dark matter halo masses for each redshift bin are given in Table~\ref{tabcorr}.
  \label{fwpz}}
\vspace{1.5ex}
\end{figure*}

The resulting $w_p(R)$ for the separate redshift bins and the power
law fits are shown in Figure~\ref{fwpz}.   For the {\qsoone}s we obtain $r_0 = 5.1\pm0.8$ \hmpc\ and $6.3\pm0.7$
\hmpc\ for the low- and high-redshift bins, respectively, and for the
{\qsotwo}s we correspondingly obtain $r_0 = 6.2\pm0.7$ \hmpc\ and
$5.8\pm1.0$ \hmpc.  The measured quasar-galaxy
cross-correlation should be largely independent between the two
redshift bins. Although the quasars are cross-correlated against the
same galaxy sample in each bin, the galaxy samples will be weighted
toward higher and lower redshifts in the high- and low-$z$ bins,
respectively.   For the high and low redshift bins, the best-fit $r_0$ values bracket
those for the full redshift samples, and are broadly consistent within
the uncertainties.  Interestingly, the best-fit clustering amplitude
for the {\qsotwo}s increases with redshift while it decreases for the {\qsotwo}s; however, given the uncertainties we decline to speculate on any
possible difference in redshift evolution between the two subsets.
Overall, the results in the different redshift bins confirm that any
variation in the observed $w_p(R)$ is sufficiently weak over the
redshift range of interest, so that the method outlined in
\S~\ref{crosscorr} should provide a reasonable estimate of the average
clustering amplitude over the full redshift range.

\subsection{Effects of quasar photo-$z$ errors}
\label{qsophotoz}

The primary uncertainty in our estimate of $w_p(R)$ for the {\qsotwo}s
is the lack of accurate (that is, spectroscopic) redshifts and
difficulty in estimating the photo-$z$ uncertainties from the neural
net calculations.  As described in \S~\ref{corranal}, in calculating
$w_p(R)$ for the {\qsotwo}--galaxy cross-correlation, we simply assume
that {\qsotwo}s lie exactly at the best redshifts output by the neural
net estimator.  Any uncertainties in the photo-$z$s or systematic
offsets from the true redshifts could therefore affect the resulting
clustering measurement.  The fact that we obtain very similar
estimates of $b_Q$ for the {\qsotwo}s from a simple angular
cross-correlation analysis as from our $w_p(R)$ calculation suggests
that uncertainties on individual photo-$z$s do not strongly affect our
estimates of the quasar bias, as long as the overall distribution in
redshifts for the {\qsotwo}s is correct.  However, it is possible that
very large discrepancies from the true photo-$z$s, or any systematic
shift in the redshift distribution, could affect both the estimate of
$w_p(R)$ and the real-space clustering parameters derived from
$\omega(\theta)$.

\begin{figure}
\epsscale{1.1}
\plotone{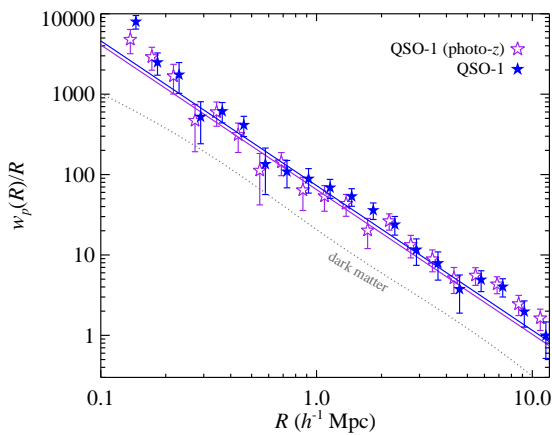}
\caption{Projected quasar-galaxy cross-correlation function for {\qsoone}s, 
using photometric redshifts (purple open stars) and
  spectroscopic redshifts (blue solid stars). Power-law fits are
  performed using a fixed $\gamma = 1.8$, and fit parameters are dark
  matter halo masses given in Table~\ref{tabcorr}.  Using the less
  accurate (photometric) redshifts has relatively small effect on
  $w_p(R)$, decreasing the best-fit power law amplitude by
  $\approx$10\%.
  \label{fagn5pz}}
\vspace{1.5ex}
\end{figure}

\begin{figure}
\epsscale{1.1}
\plotone{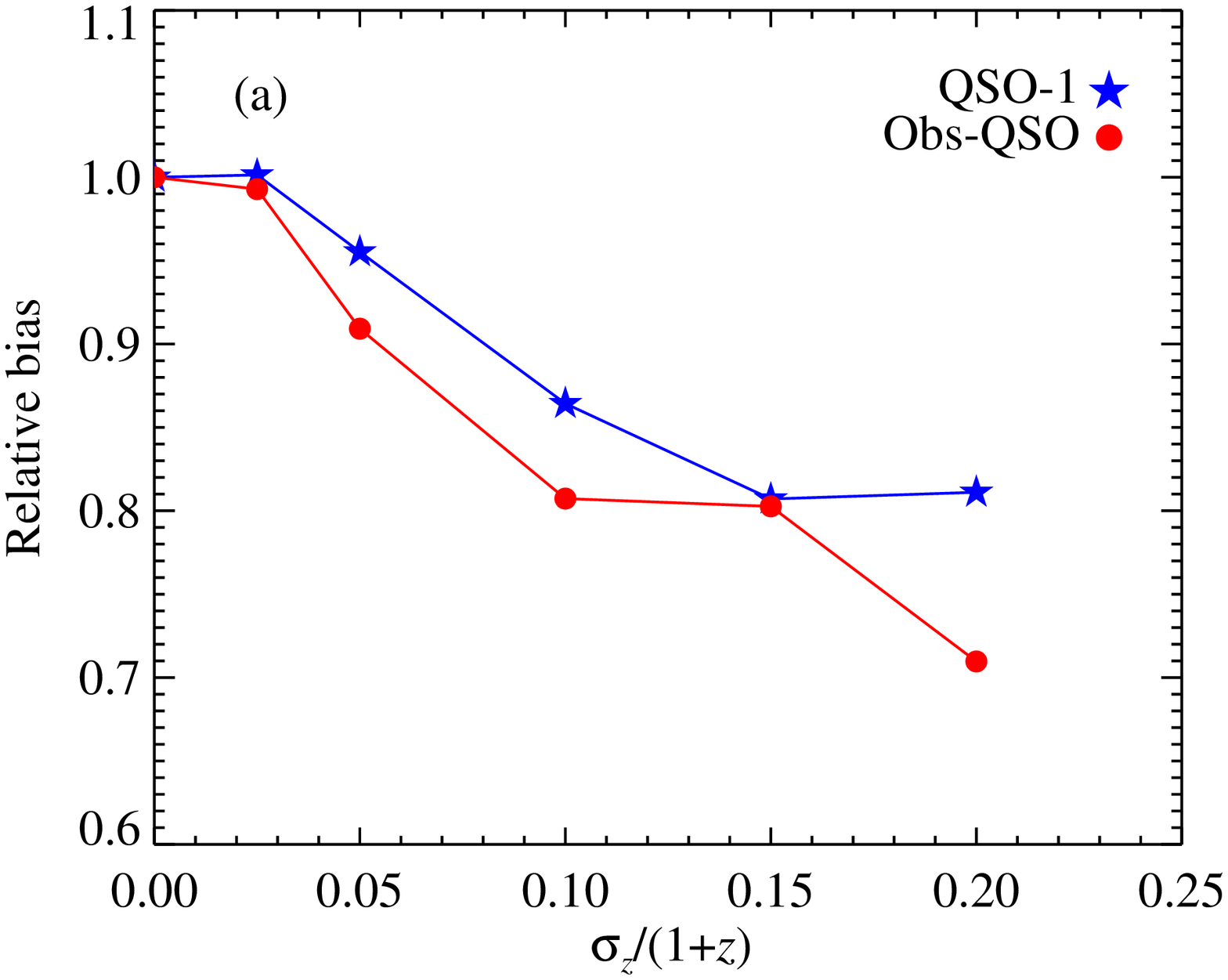}
\plotone{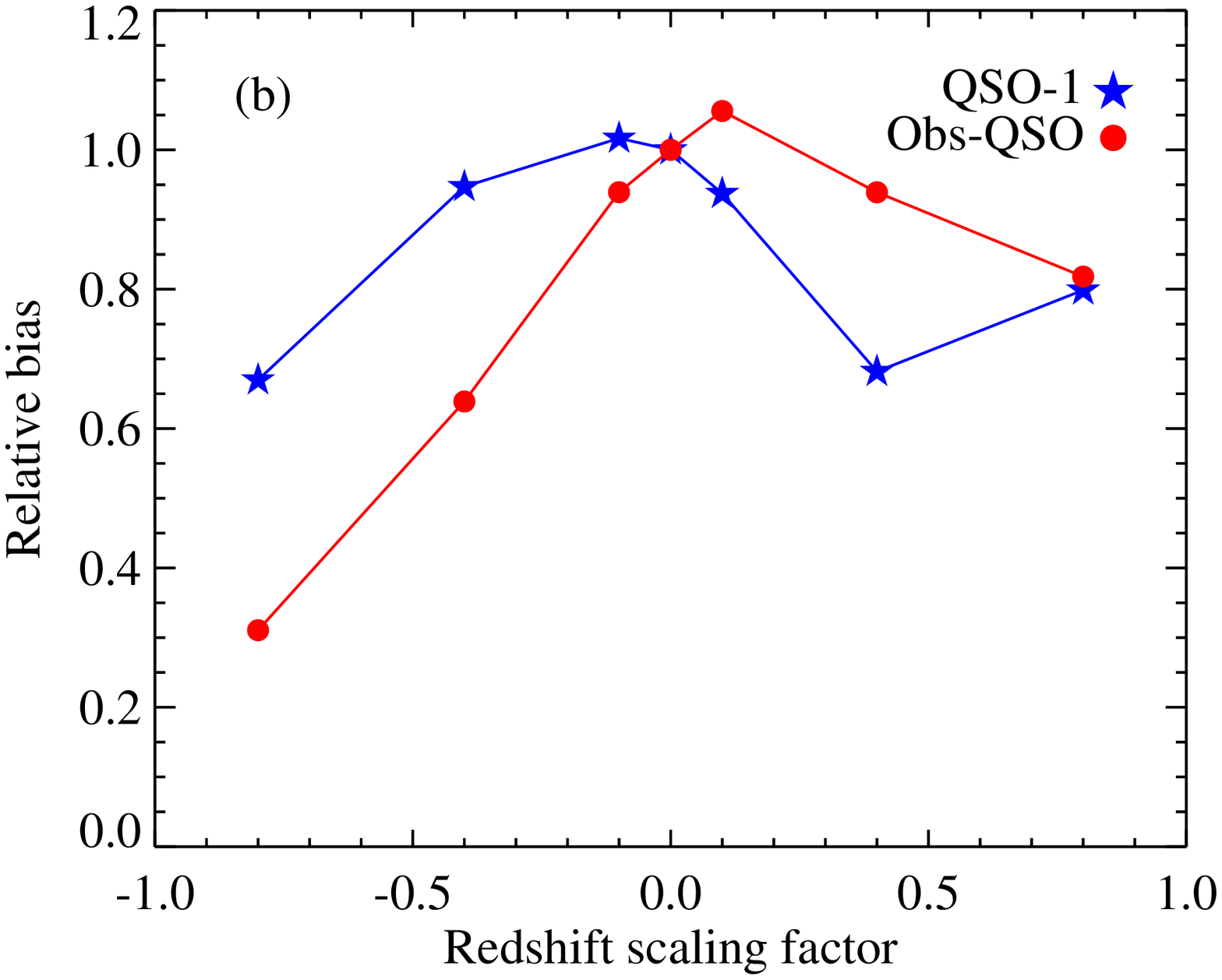}
\plotone{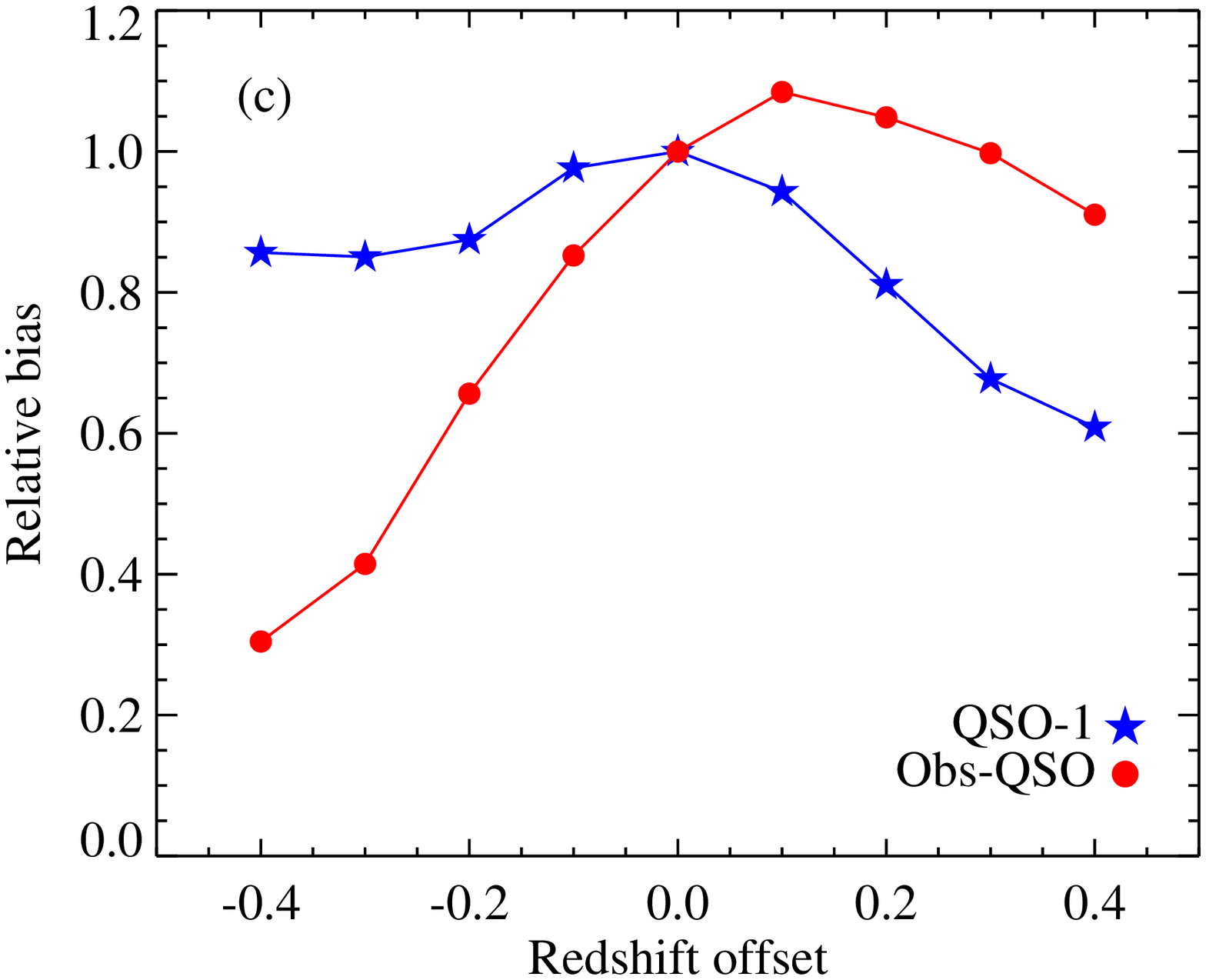}
\caption{Tests of the effects of redshift errors on the observed
  clustering, for {\qsoone}s and {\qsotwo}s, as described in \S~\ref{qsophotoz}.  (a) Relative clustering
  amplitude, after shifting the quasar redshifts by an offset drawn
  from a Gaussian random distribution with width $\sigma_z$.  Both
  types of quasar show a very similar monotonic decrease in clustering
  amplitude with $\sigma_z/(1+z)$, indicating that the photo-$z$
  estimates for the {\qsotwo}s are reasonably accurate. (b) Corresponding
  change in bias after scaling the quasar redshifts toward the limits
  of the redshift interval $0.7 < z < 1.8$; the scale parameter $S_z$
  is defined in Equation (\ref{eqnsz}). (c) Relative bias after a
  simple upward or downward shift of the quasar redshifts.  In both
  panels (b) and (c), both types of quasar show clustering amplitudes
  that peak near $S_z= 0$ or $\Delta z = 0$ indicating that there is
  no large systematic error in the photo-$z$ estimates for the {\qsotwo}s.
\label{ftest_fake_box}}
\vspace{1.5ex}
\end{figure}

To precisely explore the effect of these errors, we take advantage of
the fact that we have an equivalent sample of objects (the {\qsoone}s)
that have a similar redshift distribution and for which the redshifts
are known precisely from spectroscopy.  We can therefore adjust the
redshifts of the {\qsoone}s and re-calculate $w_p(R)$ to determine how
uncertainties or systematic shifts affect the observed correlation
amplitude.  As a simple first test, we calculate the $w_p(R)$ for the
{\qsoone}s using the photo-$z$s (as shown in Figure~\ref{fspz}) rather
than spectroscopic redshifts.  Figure~\ref{fagn5pz} shows that the
resulting $w_p(R)$ differs little from that obtained using
spectroscopic redshifts; the clustering amplitude for a power-law fit
with fixed $\gamma=1.8$ is lower by 12\%. Note that if we allow
$\gamma$ to float, the average bias for the photo-$z$ sample is
actually larger by $\approx$10\% (owing to a slightly flatter slope)
but well within the statistical uncertainties.  We conclude that for
the {\qsoone}s, photo-$z$ errors do not have a significant effect on our measurements of the clustering.

\subsubsection{Random  errors}
\label{randomerrors}
To explore photo-$z$ errors in more detail, we  systematically
test the effects of Gaussian random
errors in the quasar redshifts.  For each quasar, we shift the best
estimates of redshift (spec-$z$s for {\qsoone}s and photo-$z$s for {\qsotwo}s)
by offsets $\Delta z / (1+z)$ selected from a Gaussian random distribution with
dispersion $\sigma_z/(1+z)$.  Using these new redshifts we recalculate
$w_p(R)$, using the full formalism described in \S~\ref{corranal}.  We
perform the calculation ten times for each of several values of
$\sigma_z/(1+z)$ from 0.02 up to 0.2 (which smears the redshifts across most
of the redshift range of interest).  To ensure that this step does not
artificially smear out the redshift distribution beyond the range
probed by the galaxies, we require that the random redshifts lie
between $0.7 < z < 1.8$; any random redshift that lies outside this
range is discarded and a new redshift is selected from the 
random distribution.  For each trial we obtain the relative bias by
calculating the mean ratio of $w_p(R)$, on scales 1--10 \hmpc,
relative to the $w_p(R)$ for the best estimates of redshift.
We then average the ten trials at each $\sigma_z$ to
obtain a relation between relative bias and $\sigma_z$, shown in
Figure~\ref{ftest_fake_box}(a).

As may be expected, Figure~\ref{ftest_fake_box}({\em a}) shows that
shifting the {\qsoone} redshifts from their true values causes a
decrease in the cross-correlation amplitude, as the quasars are
preferentially correlated with galaxies that are not actually
associated in redshift space.  We find a monotonic decrease in
relative bias with $\sigma_z$, from $\approx0.95$ for $\sigma_z/(1+z)
= 0.05$ to $\approx0.8$ for $\sigma_z/(1+z) = 0.2$.  Repeating this
calculation for the {\qsotwo}s reveals a very similar trend.  The
decrease in bias with $\sigma_z/(1+z)$ shown in
Figure~\ref{ftest_fake_box}({\em a}) indicates that such errors would
affect the measurements of the clustering amplitude by at most
$\sim$20\%.

\subsubsection{Systematic shifts}
\label{systematicshifts}

While the above analysis suggests that random errors in the {\qsotwo}
photo-$z$s do not strongly affect the observed clustering amplitude,
it is also possible that systematic uncertainties in the photo-$z$ (consistent
over- or under-estimates of the redshift) could significantly alter
the observed bias.  To test this, we shift the redshifts of the
quasars as discussed in \S~\ref{randomerrors}, but in place of random
shifts, we compress all redshifts toward one end or the other of the
$0.7 < z < 1.8$ range.  (This procedure allows us to test the effects
of systematic shifts in redshift while keeping the same overall
redshift range.)  The shift in redshift is defined by a redshift
scaling parameter $S_z$, such that

\begin{eqnarray}
\label{eqnsz}
z_{\rm new} = \left\{ 
\begin{array}{l l}
  z+S_z(z-0.7) & \quad S_z < 0\\
  z+S_z(1.8-z) & \quad S_z \ge 0 \\
 \end{array} \right.
\end{eqnarray}

As an additional check we also perform a simple linear offset $\Delta
z$ of the redshifts, allowing the redshifts to move outside the
selection range of $0.7 < z < 1.8$.  As in \S~\ref{randomerrors}, we
use these new redshifts to recalculate $w_p(R)$ via the full formalism
described in \S~\ref{corranal}, and determine the relative bias on
scales 1--10 \hmpc.  Relative bias versus $S_z$ and $\Delta z$ are
shown in Figures~\ref{ftest_fake_box}(b) and (c).  For the {\qsoone}s, the
peak of the observed clustering amplitude is very close to $S_z = 0$,
while shifting the redshifts down or up systematically decreases the
bias.  The {\qsotwo}s show a similar peak near $S_z = 0$, indicating that
the {\qsotwo} photo-$z$s are not systematically offset higher or lower
than the true redshifts by a large factor.  We note that for the {\qsotwo}s a slight shift to higher redshifts ($\Delta z = 0.1$) increases the
clustering by a small amount ($\approx$8\%).

Finally, we emphasize that any possible low-redshift contaminants
\citepalias[such as star-forming galaxies, as discussed in \S~7
  of][]{hick07abs}, will serve only to decrease the observed
clustering signal, as they will be completely uncorrelated in angular
space with the higher-redshift SDWFS galaxies that lie in entirely
separate large-scale structures.
Therefore the observed $w_p(R)$ represents a robust lower limit to the
clustering amplitude for the {\qsotwo}s.

\begin{figure*}
\epsscale{1.15}
\plottwo{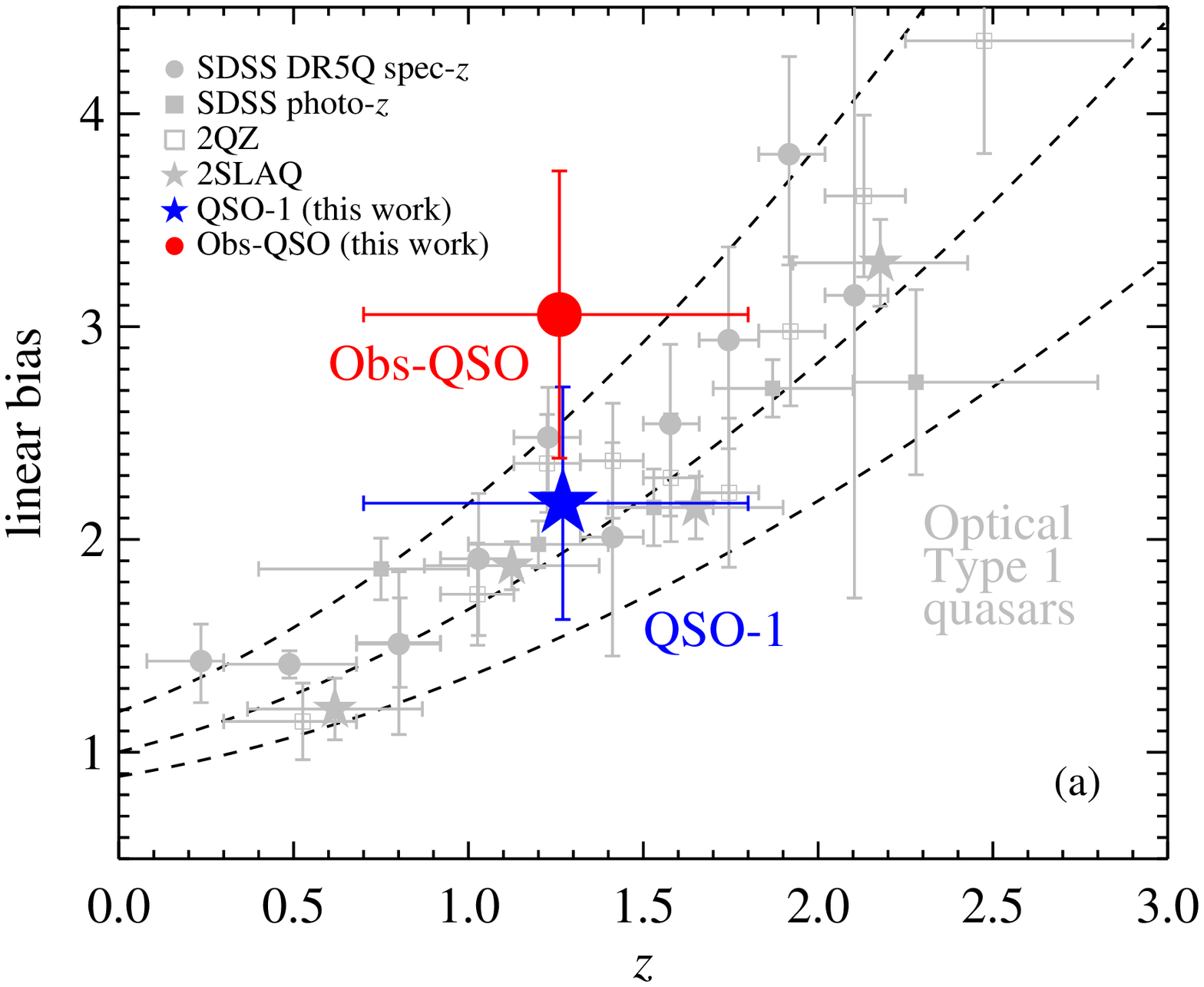}{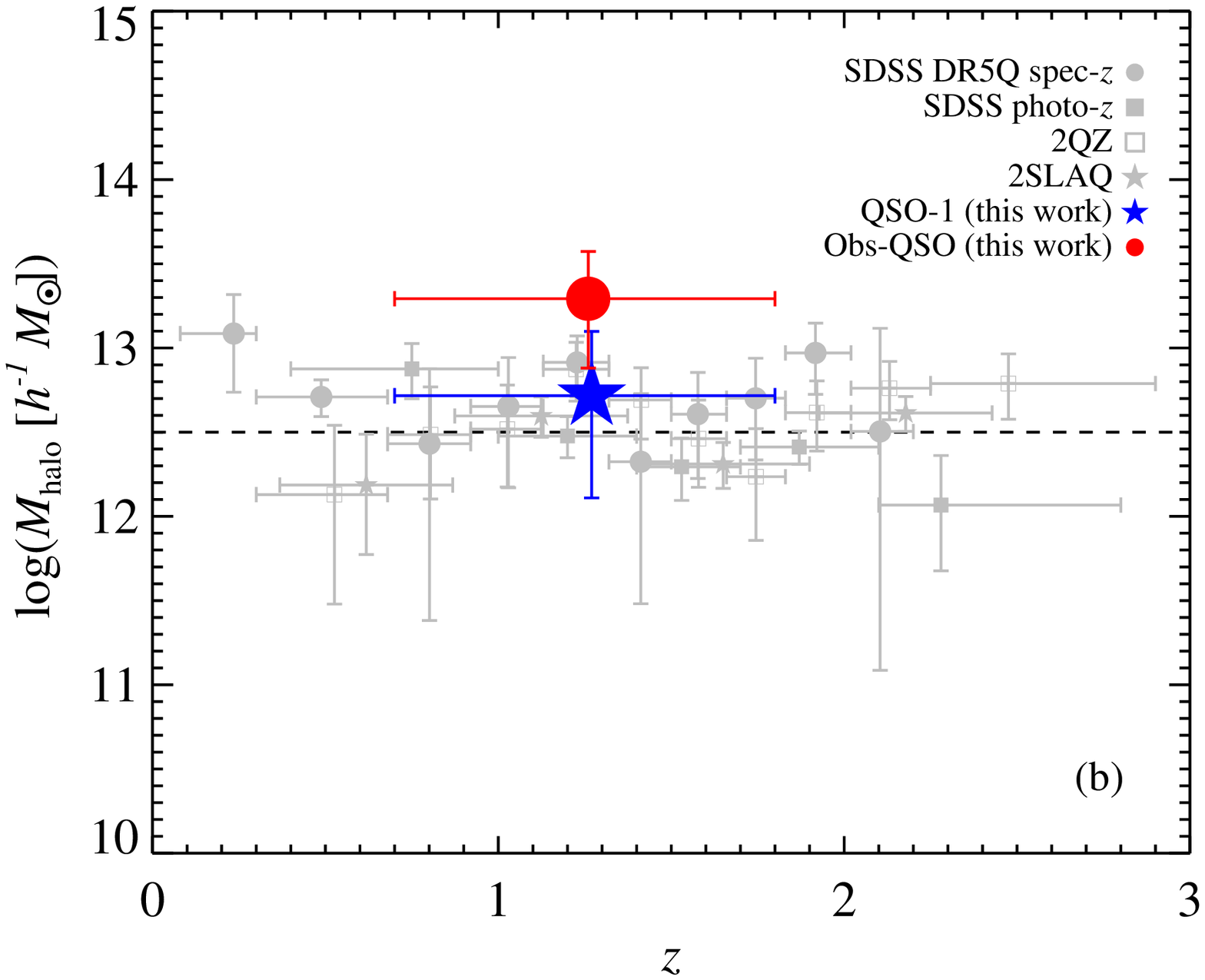}
\caption{Comparison of (a) linear bias and (b) inferred dark matter
  halo mass to previous measurements of Type 1 quasar
  clustering.  Lines in the left plot correspond to the linear bias for dark matter halos with $\log{M_{\rm halo} [h^{-1} M_{\sun}]} = 12.0$, 12.5, and 13.0 (from bottom to top).  The gray points show results from quasar-quasar
  correlation measurements using spectroscopic samples from SDSS
  \citep{ross09qsoclust}, 2QZ \citep{croo05}, and 2SLAQ
  \citep{daan08clust}, as well as the clustering of
  photometrically-selected quasars from SDSS \citep{myer06clust}.  The
  blue star and red circle show our measurements for {\qsoone}s and {\qsotwo}s,
  respectively.  Our {\qsoone} measurement agrees closely with previous
  work on Type 1 quasar-quasar correlations, while the {\qsotwo}s show
  marginally $\approx 1\sigma$ stronger clustering, corresponding to a factor of roughly four in $M_{\rm halo}$.
\label{fcompare}}
\vspace{1.5ex}
\end{figure*}

\section{Discussion}

\label{discussion}

We have used the IR-selected quasar sample of \citetalias{hick07abs}
to measure the clustering amplitude and to estimate characteristic
dark matter halo masses for roughly equivalent samples of unobscured and obscured quasars.  We obtain highly
significant detections of the clustering for both samples, with
marginally stronger clustering for the {\qsotwo}s.  In this section, we
compare our results for {\qsoone}s to previous results on unobscured
quasar clustering, speculate on physical explanations for possible
stronger clustering for {\qsotwo}s, and discuss prospects for future
studies with the next generation of observatories.

\subsection{Comparison with previous results}

We compare our observed absolute bias for the mid-IR selected
quasars to the clustering of optically-selected (Type 1) quasars,
which has been well established by a number of works.  Among the most
precise measurements to date are studies that have used data from the
2dF and SDSS surveys.  Recently, \citet{ross09qsoclust} measured the
evolution of the quasar 3-D autocorrelation function based on
spectroscopic quasars from SDSS Data Release 5, and compared to
previous results from spectroscopic samples from the 2QZ
\citep{croo05} and 2SLAQ \citep{daan08clust}, as well as the
clustering of photometrically-selected quasars from SDSS
\citep{myer06clust}.  Figure~\ref{fcompare}(a) shows the redshift
evolution of linear bias for Type 1 quasars from these studies, taken
from Figure~12 of \citet{ross09qsoclust}.  Where appropriate, we
have converted the bias to our adopted cosmology using the formalism
in the Appendix of \citet{star10xclust_aph}, assuming a shape factor
$s=1$ for simplicity.  Figure~\ref{fcompare}(b) shows the
corresponding estimates of characteristic $M_{\rm halo}$ derived from
the linear bias using the prescription of \citet{shet01halo}.  The
linear bias and halo masses for the {\qsoone}s and {\qsotwo}s are
shown for comparison.

It is readily apparent from Figure~\ref{fcompare}(a) that the observed
bias of Type 1 quasars increases with redshift, as discussed in
\S~\ref{intro}.  (These results are also consistent with a number of
other quasar clustering studies using other data see e.g., Figure~15
of \citealt{hopk08frame1}.)  The dashed curves in
Figure~\ref{fcompare}(a) show the increase in bias with redshift for
halos of constant  mass, clearly showing that at all redshifts the
{\qsoone}s reside in dark matter halos of roughly a few
$\times10^{12}$ \hmsun.  Our measurement for the {\qsoone}s is in
excellent agreement with the evolution in linear bias and roughly
constant $M_{\rm halo}$ observed in previous measurements of Type 1
quasar clustering.

For the {\qsotwo}s, the best-fit bias is marginally larger ($\approx 1
\sigma$), corresponding to a factor of roughly four difference in the
characteristic $M_{\rm halo}$ between the {\qsoone}s and {\qsotwo}s
(Figure~\ref{fcompare}b).  As discussed above, random errors in the
photo-$z$s can only decrease the observed clustering amplitude and the
inferred $M_{\rm halo}$.  Thus our measurement of {\qsotwo} clustering
represents a lower limit, and it is possible that the true {\qsotwo}
bias is somewhat higher (although the results of \S~\ref{verification}
suggest the true bias may be larger by at most $\sim$20\%).  Based on
this analysis we can make the robust conclusion that the {\qsotwo}s
are {\em at least} as strongly clustered as their {\qsoone}
counterparts.

\subsection{Physical implications}

Stronger clustering for obscured quasars would have significant
implications for physical models of the obscured quasar population.
In terms of unified models, a difference in clustering between
obscured and unobscured quasars would rule out the simplest picture in
which obscuration is purely an orientation effect, but may be
consistent with more complicated scenarios where the effective
covering fraction changes with environment.  Alternatively, if
obscuration is caused by large ($\sim$kpc scale) structures, then the
processes that drive these asymmetries (e.g., mergers, disk
instabilities, accretion of cold gas from the surrounding halo) may be
more common in halos of larger mass.  Indeed, given that some fraction
of quasars might naturally be expected to be obscured by a
``unified''-model torus, any observed differences in clustering may
reflect even stronger intrinsic dependence of large-scale obscuration
and environment.

An intriguing scenario for obscured quasars is that they represent an
early evolutionary phase of rapid black hole growth before a
``blowout'' of the obscuring material from the central regions of the
galaxy and the emergence of an unobscured quasar \citep[e.g., Figure~1
of][]{hopk08frame1}.  Quasars tend to radiate at large fraction of the
Eddington rate (\citealt{mclu04mbh, koll06}; although see
\citealt{kell10qsoedd}), so that the similar $L_{\rm bol}$ for
{\qsoone}s and {\qsotwo}s would imply that they host black holes of
similar masses. Any correlation \citep[e.g.,][]{ferr02mbhbulge,
  boot10mhalo}, even if indirect \citep[e.g.,][]{korm11mbhbulge},
between the final masses of black holes and those of their host halos
would thus suggest that our obscured and unobscured quasars would have
the same $M_{\rm halo}$, as long as their black holes are near their final
masses. However, if obscured quasars are in an earlier phase of rapid
growth and so are in the process of ``catching up'' to their final
mass \citep[e.g.,][]{king10bhmass}, then they would have a larger
$M_{\rm halo}$ compared to unobscured quasars with the same $M_{\rm
  BH}$.  In light of recent debate as to whether black holes generally
grow before or after their hosts \citep[e.g.,][]{peng06b,
  alex08bhmass, woo08qsombh, deca10mbhb}, this scenario would imply
that black hole growth lags behind that of the host halo.

In any physical picture, a significant difference in clustering
between obscured and unobscured quasars would also imply a difference
in accretion duty cycles (or equivalently, lifetime).  {\qsoone}s and
{\qsotwo}s are found in roughly equal numbers, but the abundance of
dark matter halos drops rapidly with mass \citep[e.g.,][]{jenk01halo},
thus implying that if one type of quasars are found in larger halos
then they must be longer-lived.  With our current results, we are able
to rule out any model in which obscured quasars are substantially {\em
  less} strongly clustered or have {\em shorter} lifetimes than their
unobscured counterparts. With more accurate future measurements,
detailed studies of halo masses and lifetimes for obscured and
unobscured quasars could place powerful constraints on evolutionary
scenarios such as those described above.

\subsection{Future prospects}

Our results demonstrate the potential for studying the clustering of
obscured quasars in extragalactic multiwavelength surveys, and the
marginally significant difference in clustering we observe for
obscured and unobscured quasars provides strong motivation for more
precise measurements in the future.  The two main avenues for progress
are improvements in redshift accuracy and selection of larger samples
for better statistical accuracy.  Upcoming sensitive, wide-field
multi-object spectrographs will enable efficient measurements of
redshift for large numbers of optically-faint sources and so improve
calibrations of obscured quasar photo-$z$s, or with large enough
samples, enable fully 3-D clustering studies.  In addition, we will
soon have the capability to detect many thousands of obscured quasars
based on very wide-field observations in the mid-IR with the {\em
  Wide-Field Infrared Survey Explorer} \citep{wrig08wise} and in
X-rays with eROSITA \citep{pred07erosita} or the {\em Wide-Field X-ray
  Telescope} \citep{murr10wfxt}.  These data sets will allow us to measure
obscured quasar clustering with statistical precision that is
comparable to current measurements of unobscured quasars.

\section{Summary}

We have used data from the \bootes\ wide-field multiwavelength survey
to measure the two-point spatial cross-correlation between unobscured
({\qsoone}) and obscured ({\qsotwo}) mid-IR selected quasars in the redshift
range $0.7 < z < 1.8$.  The {\qsoone}s exhibit clustering corresponding to
a typical $M_{\rm halo}\sim 5 \times 10^{12}$ \hmsun, similar to
previous studies of optically-selected quasar clustering.  We
robustly determine that the {\qsotwo}s are clustered {\em at least} as
strongly as the {\qsoone}s, with a marginally stronger signal
corresponding to host halos of mass $\sim 2\times10^{13}$ \hmsun; the
true clustering amplitude could be up to $\sim$20\% larger owing to photo-$z$
uncertainties for the {\qsotwo}s that can decrease the observed
correlation amplitude.  Our results motivate more accurate
measurements of obscured quasar clustering with larger quasar samples
and more accurate redshifts.  If future studies confirm that obscured
quasars are more strongly clustered than their their unobscured
counterparts, this would rule out the simplest ``unified'' models and
may provide evidence for scenarios in which rapid obscured accretion
represents an evolutionary phase in the growth of galaxies and their
central black holes.

\begin{acknowledgements}
  We thank our colleagues on the NDWFS, AGES, SDWFS, and XBo\"otes
  teams.  We thank the anonymous referee for helpful comments that
  improved the paper, and Philip Hopkins and Peder Norberg for
  productive discussions.  The NOAO Deep Wide-field Survey, and the
  research of A.D. and B.T.J. are supported by NOAO, which is operated
  by the Association of Universities for Research in Astronomy (AURA),
  Inc.\ under a cooperative agreement with the National Science
  Foundation.  This paper would not have been possible without the
  efforts of the \chandra, \spitzer, KPNO, and MMT support staff.
  Optical spectroscopy discussed in this paper was obtained at the MMT
  Observatory, a joint facility of the Smithsonian Institution and the
  University of Arizona.  The first {\em Spitzer} MIPS survey of the
  \bootes\ region was obtained using GTO time provided by the {\em
    Spitzer} Infrared Spectrograph Team (PI: James Houck) and by
  M. Rieke.  We thank the collaborators in that work for access to the
  24 micron catalog generated from those data.  R.C.H.\ was supported
  by an STFC Postdoctoral Fellowship and an SAO Postdoctoral
  Fellowship, and A.D.M.\ was generously funded by the NASA ADAP
  program under grant NNX08AJ28G.  D.M.A.\ is grateful to the Royal
  Society and Philip Leverhulme Prize for their generous support.
  R.J.A.\ was supported by the NASA Postdoctoral Program, administered
  by Oak Ridge Associated Universities through a contract with NASA.
\end{acknowledgements}

\appendix

\section{Edge effects and correlations on large scales}

\begin{figure}
\epsscale{0.6}
\plotone{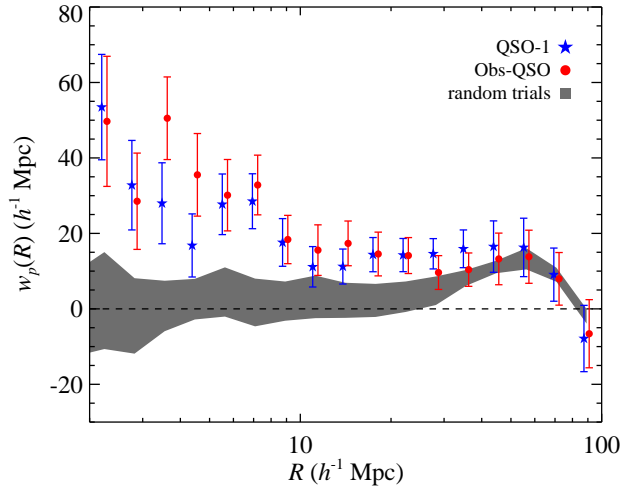}
\vspace{-1.5ex}
\caption{\label{fwprand} Observed projected cross-correlation between
  quasars and galaxies derived using Eqn.~(\ref{eqnwp}) (blue stars
  and red circles for {\qsoone}s and {\qsotwo}s, respectively), compared to the same
  quantity but derived for samples of quasars with the redshifts equal
  to those of the {\qsoone}s but with randomized sky positions.  The gray
  shaded region shows the dispersion between ten different random
  quasar samples.  Note that the correlation between random quasars
  and galaxies is small relative to the real-quasar galaxy
  correlation, except on scales $>$10--20 \hmpc\ where both quantities
  show a hump and then become negative.  This feature is likely due to
  edge effects in the finite SDWFS field, as discussed in the
  Appendix.  Because of this artifact we restrict our correlation
  analyses to $R < 12$ \hmpc.  }
\end{figure}

The large area of the \bootes\ survey allows us to measure galaxies at
relatively large physical separations; 1 degree corresponds to 50
\hmpc\ at $z=1.5$.  However, when we use Equation (\ref{eqnwp}) to
calculate the projected real-space correlation function on large
scales, we find that the $w_p(R)$ flattens on scales $R\sim10$--20
\hmpc, corresponding to tens of arcmin, and then becomes negative at
$R\sim 100$ \hmpc\ (Figure~\ref{fwprand}).  This behavior is not
observed for quasar clustering from other, wider-field surveys
\citep[e.g.][]{croo05, myer09clust}, for which the correlation
function continues to decrease on larger scales.  While the integral
constraint require that the correlation function becomes negative on
some scales, in galaxy auto-correlation surveys this generally only
happens at $R\gtrsim$ 200 \hmpc\ \citep[e.g.,][]{eise05bao}.

One possibility is that the observed behavior is due edge effects
arising from the finite geometry of the SDWFS field, which are not
taken into account by the simple $\xi(R) = DD/DR - 1$ estimator in the
\citetalias{myer09clust} formalism \citep[e.g.,][]{land93}.  To test
this possibility, we re-performed the correlation analysis described
in \S~\ref{corranal}, after randomizing the positions of the quasars
on the sky within the area of good SDWFS photometry.  We performed 10
separate random trials, for which the cross-correlations are shown in
Figure~\ref{fwprand} along with the $w_p(R)$ values for the {\qsoone}s and
{\qsotwo}s.  It is clear from Figure~\ref{fwprand} that on scales $\lesssim$10
\hmpc, the projected cross-correlation between the random quasars and
galaxies is small compared to the $w_p(R)$ for the real quasar sample.
However, on scales $\gtrsim$20 \hmpc, both the real and random samples
show an increase in $w_p(R)$ which eventually becomes negative around
$R=100$ \hmpc.  The quantities of interest in this paper
(i.e. absolute bias and dark matter halo mass) can be measured by
studying the correlations on scales $<$12 \hmpc, where the artifacts
are small and have negligible impact on the fits to the correlation
function.  For this paper we therefore limit the correlation analyses
to those scales.

\end{document}